\documentclass[12pt,draftclsnofoot,onecolumn]{IEEEtran}

\usepackage{array,graphicx,cite,epsfig,subfigure,amssymb,amsfonts}
\usepackage{bbm,amsmath,epstopdf,algorithm,algorithmic, multirow}
\usepackage{comment}
\usepackage{color}
\usepackage{cases}
\usepackage{slashbox}
\usepackage{pifont} 

\newcommand{\cmark}{\ding{51}}
\newcommand{\xmark}{\ding{55}}

\newcommand{\be}{\begin{equation}}
\newcommand{\ee}{\end{equation}}
\def\bee#1\eee{\begin{align}#1\end{align}}
\newcommand{\bse}{\begin{subequations}}
\newcommand{\ese}{\end{subequations}}
\newcommand{\nnb}{\nonumber}

\newtheorem{theorem}{\textbf{Theorem}}

\newtheorem{lemma}{\textbf{Lemma}}

\newtheorem{conjecture}{\textbf{Conjecture}}

\newcommand{\specialcell}[2][c]{%
  \begin{tabular}[#1]{@{}c@{}}#2\end{tabular}}

\newcommand{\red}[1]{{\color{red}#1}}
\newcommand{\blue}[1]{{\color{blue}#1}}

\newcommand{\bm}[1]{\boldsymbol{#1}}

\def \ISTR {}

\begin{document}

\title{The Story of $1/e$: ALOHA-based and Reinforcement-Learning-based Random Access for Delay-Constrained Communications}

\author{Lei~Deng,~\IEEEmembership{Member,~IEEE,}
        Danzhou~Wu,
        Jing~Deng,~\IEEEmembership{Fellow,~IEEE,}
        Po-Ning~Chen,~\IEEEmembership{Senior Member,~IEEE,}
        and Yunghsiang~S.~Han,~\IEEEmembership{Fellow,~IEEE}
\thanks{L.~Deng and D.~Wu are with
College of Electronics and Information Engineering,
Shenzhen University (email: ldeng.sjtu@gmail.com, wudanzhou2019@email.szu.edu.cn).}
\thanks{J.~Deng is with Department of Computer Science, University of North Carolina at Greensboro (email: jing.deng@uncg.edu).}
\thanks{P.-N.~Chen is with Department of Electrical and Computer
Engineering, National Yang Ming Chiao Tung University (email: poningchen@nycu.edu.tw).}
\thanks{ Y.~S.~Han (corresponding author) is with
Shenzhen Institute for Advanced Study, University
 of Electronic Science and Technology of China (email: yunghsiangh@gmail.com).}
\thanks{Part of this work was invited to be presented in IEEE ICCCN 2018 \cite{dengon2018}.}}

\maketitle

\begin{abstract}

Motivated by the proliferation of real-time applications in multimedia communication systems, tactile Internet, and cyber-physical systems, supporting delay-constrained traffic becomes critical for such systems. In delay-constrained traffic, each packet has a hard deadline; when it is not delivered before its deadline is up, it becomes useless and will be removed from the system. In this work, we focus on designing random access schemes for delay-constrained wireless communications. We first investigate three ALOHA-based schemes and prove that the system timely throughput of all three schemes under corresponding optimal transmission probabilities asymptotically converges to $1/e$, same as the well-known throughput limit for delay-unconstrained ALOHA systems. The fundamental reason why ALOHA-based schemes cannot achieve asymptotical system timely throughput beyond $1/e$ is that all active ALOHA stations access the channel with the same probability in any slot. To go beyond $1/e$, we propose a reinforcement-learning-based scheme for delay-constrained wireless communications, called RLRA-DC, under which different stations collaboratively attain different transmission probabilities by only interacting with the access point. Our numerical result shows that the system timely throughput of RLRA-DC can be as high as 0.8 for tens of stations and can still reach 0.6 even for thousands of stations, much larger than $1/e$.

\end{abstract}


\section{Introduction}
\IEEEPARstart{W}ireless communication is shifting its role from connecting people to network everything in multiple vertical domains,
many of which have hard delay constraint. Typical examples include multimedia communication systems such
as real-time streaming and video conferencing \cite{deng2017timely}, tactile Internet \cite{fettweis2014tactile,simsek20165g},
networked control systems (NCSs) such as remote control of unmanned aerial vehicles (UAVs) \cite{baillieul2007control,zeyu2016autonomous},
and cyber-physical systems (CPSs) such as medical tele-operations, X-by-wire vehicles/avionics, factory automation, and robotic collaboration \cite{kang2013design}.
In such applications, each packet has a hard deadline: if it is not delivered before the deadline, it becomes useless and will be removed from the system.
As an indispensable part of communication systems, how to design efficient random access schemes to support such delay-constrained applications
is important but introduces new challenges \cite{popovski2018wireless,Hsieh2018Decentralized}. In this work, we will investigate
both traditional ALOHA-based random access schemes and model learning-based random access schemes for delay-constrained wireless communications.

Since  Abramson's invention in 1970 \cite{abramson1970aloha}, ALOHA-type protocols have been widely used for multiple users to access a shared communication channel
due to its extreme simplicity and decentralized nature. One popular type is \emph{slotted ALOHA}, where users are synchronized and can
only transmit at the beginning of a slot \cite{roberts1975aloha}.
It is well-known that the optimal asymptotic system throughput of slotted ALOHA system is $1/e=36.8\%$ \cite{kurose2013computer}.
In addition, there are many works to investigate the stability region of slotted ALOHA, i.e.,
to characterize the feasible input rates under which the system can be stabilized in the sense
that {the size of all queues will not diverge to infinity} \cite{tsybakov1979ergodicity,rosenkrantz1983instability,hajek1982decentralized,ghez1988stability,rao1988stability,szpankowski1994stability}.
There are also many types of extension for slotted ALOHA protocol,
including multi-packet reception\cite{naware2005stability,ghez1988stability}, framed slotted ALOHA \cite{lee2005enhanced,prodanoff2010optimal,yu2017stability}, coded slotted ALOHA \cite{paolini2015coded}, slotted ALOHA with successive interference cancellation (SIC) \cite{casini2007contention}, etc. These works focus on delay-unconstrained case in the sense that any packet can be delivered in however much time.

There have been a few  works on slotted ALOHA with delay-constrained traffic.
Birk et al. in \cite{birk1999judicious} proposed a scheme to increase the capacity
by dividing packets into multiple sub-packets
and transmitting redundant coded copies of the sub-packets.
Stefanovic et al. in \cite{stefanovic2017frameless} studied delay-constrained frameless ALOHA
in the finite-length case.
Malak et al. in \cite{malak2017throughput} investigated how to maximize the throughput of a random access system
subject to given constraints on latency and outage.
Borst and Zubeldia in \cite{borst2018delay} studied the scaling result of
the average delay of slotted ALOHA systems in terms of the number of stations.
Zhang et al. in \cite{zhang2017optimal} investigated the system throughput
and the optimal retransmission probability with delay-constrained \emph{saturated} traffic
in the sense that each station always has a new packet arrival once its queue becomes empty (namely
each station always has a packet to transmit). We should note that such a saturated traffic model is
rather  impractical. For example, in NCSs and CPSs, the control messages usually arrive periodically.
Thus, in this paper we analyze the slotted ALOHA system
with a non-saturated delay-constrained traffic model.
It is called frame-synchronized traffic patten, which was widely investigated in the packet scheduling policy design in
the delay-constrained wireless communication community \cite{hou2009qos,deng2017timely}.

In addition to traditional ALOHA-based random access schemes, recently there are also a few works on learning-based access protocol.
Deep reinforcement learning (DRL) has been introduced into the random access scheme design in \cite{yiding2019deep} and \cite{yiding2020non}.
In \cite{yiding2019deep}, Yu et al. proposed a scheme, called deep-reinforcement learning multiple access (DLMA), adopted feedforward neural networks (FNN) as the deep neural network.
\cite{yiding2020non} applied DRL into CSMA and designed a new CSMA variant, called CS-DLMA.
Both \cite{yiding2019deep} and \cite{yiding2020non} assume a saturated delay-unconstrained traffic pattern.
Wu et al. designed an R-learning-based random access scheme in a two-user delay-constrained heterogeneous wireless network in \cite{danzhou2021reinforcement}.
Their proposed scheme, called tiny state-space R-learning random access (TSRA),
achieves higher timely throughput, lower computation complexity than DLMA.
In addition, \cite{destounis2019learn2mac} proposed an online-learning-based access scheme, called Learn2MAC, to provide delay guarantee and low energy consumption.

In this paper,  centering around the well-known throughput limit $1/e$, we will first study three different delay-constrained slotted ALOHA schemes and then propose a reinforcement-learning-based random access scheme.
The three ALOHA-based schemes are \emph{$p$-constant slotted ALOHA} where the retransmission probability
is a constant all the time, \emph{$p$-dynamic slotted ALOHA} where the
retransmission probability changes according to the number of active stations at any slot,
and \emph{framed slotted ALOHA} where each station randomly selects a slot in a frame to probabilistically transmit its packet.
Since slotted ALOHA is a popular wireless access scheme, investigating the
fundamental performance of slotted ALOHA protocol to deliver delay-constrained traffic
will add understandings and shed light on the design of practical wireless access protocols.
We theoretically prove that the system timely throughput of all three ALOHA-based schemes converges to $1/e$ as the number of stations goes to infinity,
as summarized in Table \ref{tab:main-results}.
The fundamental reason why ALOHA-based schemes cannot achieve asymptotical system timely throughput beyond $1/e$
is that all active ALOHA stations access the channel with the same probability in any slot.
Since the number of active stations tends to be infinite as the number of total stations goes to infinity, the system timely throughput also converges to $1/e$.
In order to go beyond $1/e$, we propose a Reinforcement-Learning-based Random Access scheme for Delay-Constrained communications, called RLRA-DC, which can effectively reduce the competition level among active stations and improve the system timely throughput. RLRA-DC is designed based on R-learning \cite{schwartz1993reinforcement, singh1994reinforcement,sutton2018reinforcement},
a less-popular variant of reinforcement learning different from the widely used Q-learning.

In particular, our contributions of this paper are as follows:
\begin{itemize}
\item We prove that the maximum system timely throughput converges to $1/e=36.8\%$ as the number of stations $N$ goes to infinity
for any hard delay $D$ for all three ALOHA-based schemes. This delay-constrained asymptotic result
is the same as the asymptotic maximum system throughput for delay-unconstrained slotted ALOHA system with saturated traffic \cite[Chapter 5.3.2]{kurose2013computer}.
For $p$-constant slotted ALOHA, we prove that the optimal retransmission probability behaves asymptotically as $1/N$.
\item In the finite regime of $N$,
we propose an algorithm with time complexity $O( \min\{D,N\} \cdot D)$
to compute the system timely throughput for $p$-constant slotted ALOHA (resp. $p$-dynamic slotted ALOHA)
for any number of stations $N$, and any hard delay $D$, and any retransmission probability $p$ (resp. any retransmission policy $\bm{p}$).
For $p$-dynamic slotted ALOHA, we further derive the closed-form optimal dynamic retransmission policy.
\item For framed slotted ALOHA with finite $N$, we derive the closed-form optimal retransmission probability and the
closed-form maximum system timely throughput.
\item
To go beyond $1/e$, we introduce reinforcement learning to design a novel random access scheme, called RLRA-DC, which can effectively reduce competition among active stations.
Under our proposed random access scheme, stations in the system cooperate with each other to intelligently reduce the competition level to improve the system timely throughput.
\item We conduct extensive simulations to confirm the correctness of our theoretical analysis for three ALOHA-based schemes.
We also show that the system timely throughput of thousands of stations when our proposed RLRA-DC scheme is deployed reaches 0.6, which is much higher than $1/e$ of the ALOHA-based systems.
\end{itemize}

\begin{table*}
\centering
\caption{Main results on three different delay-constrained slotted ALOHA schemes. Here $D$ is the hard delay of all packets,
$N$ is the total number of stations, and $n(t)$ is the number of active stations at slot $t$.}
\label{tab:main-results}
\begin{tabular}{|c|c|c|c|c|c|}
\hline
\specialcell{Scheme of \\ Slotted ALOHA}                     & System Timely Throughput                                             & Retransmission Probability                 & Feedback  & $N$      & $n(t)$  \\ \hline
$p$-constant  & $\lim\limits_{N \to \infty} R^*(D,N)=\frac{1}{e}$                    & $\lim\limits_{N\to\infty} N \cdot p^*(D,N)=1$     & \cmark    & \cmark   & \xmark  \\ \hline
$p$-dynamic & $\lim\limits_{N \to \infty} R_{\textsf{dynamic}}^*(D,N)=\frac{1}{e}$ & $p_{\textsf{dynamic}}^*(t)=\frac{1}{n(t)}$ & \cmark    & \cmark   & \cmark  \\ \hline
Framed   & \hspace{-0.2cm} \specialcell{$R_{\textsf{framed}}^*(D,N)=
\left\{
  \begin{array}{ll}
    \hspace{-0.2cm} \frac{N}{D-1} \cdot \left(\frac{D-1}{D}\right)^N, & \hspace{-0.3cm} \hbox{if $N \le D$;} \\
    \hspace{-0.2cm} \left(1-\frac{1}{N}\right)^{N-1}, & \hspace{-0.3cm} \hbox{if $N > D$.}
  \end{array}
\right.
$
\hspace{-0.2cm}
\\$\lim\limits_{N \to \infty} R_{\textsf{framed}}^*(D,N)=\frac{1}{e}$}  & \hspace{-0.2cm} {$p_{\textsf{framed}}^*(D,N) = \min\left\{\frac{D}{N},1\right\}$} & \xmark    & \cmark   & \xmark  \\ \hline
\end{tabular}
\end{table*}

The rest of the paper is organized as follows. We describe our system model in Section~\ref{sec:model}.
For $p$-constant slotted ALOHA, we propose a polynomial-time algorithm
to compute the system timely throughput in Section~\ref{sec:algorithm} and
analyze its asymptotic performance in Section~\ref{sec:asymptotic-performance}.
In Section~\ref{sec:two-variants}, we analyze $p$-dynamic slotted ALOHA and framed slotted ALOHA.
In Section~\ref{sec:rl-algorithm}, we propose the reinforcement-learning-based random access scheme RLRA-DC.
We use simulation to validate our theoretical analysis and also demonstrate the beyond-$1/e$ performance of RLRA-DC.
Section~\ref{sec:conclusion} concludes this paper.

\ifx \ISTR \undefined
Due to the page limit, we move part of technical proofs to our technical report \cite{TR}, without which this paper remains self-contained and includes major technical contributions and contents.
\else
\fi

\section{System Model} \label{sec:model}
\begin{figure}
  \centering
  \includegraphics[width=0.6\linewidth]{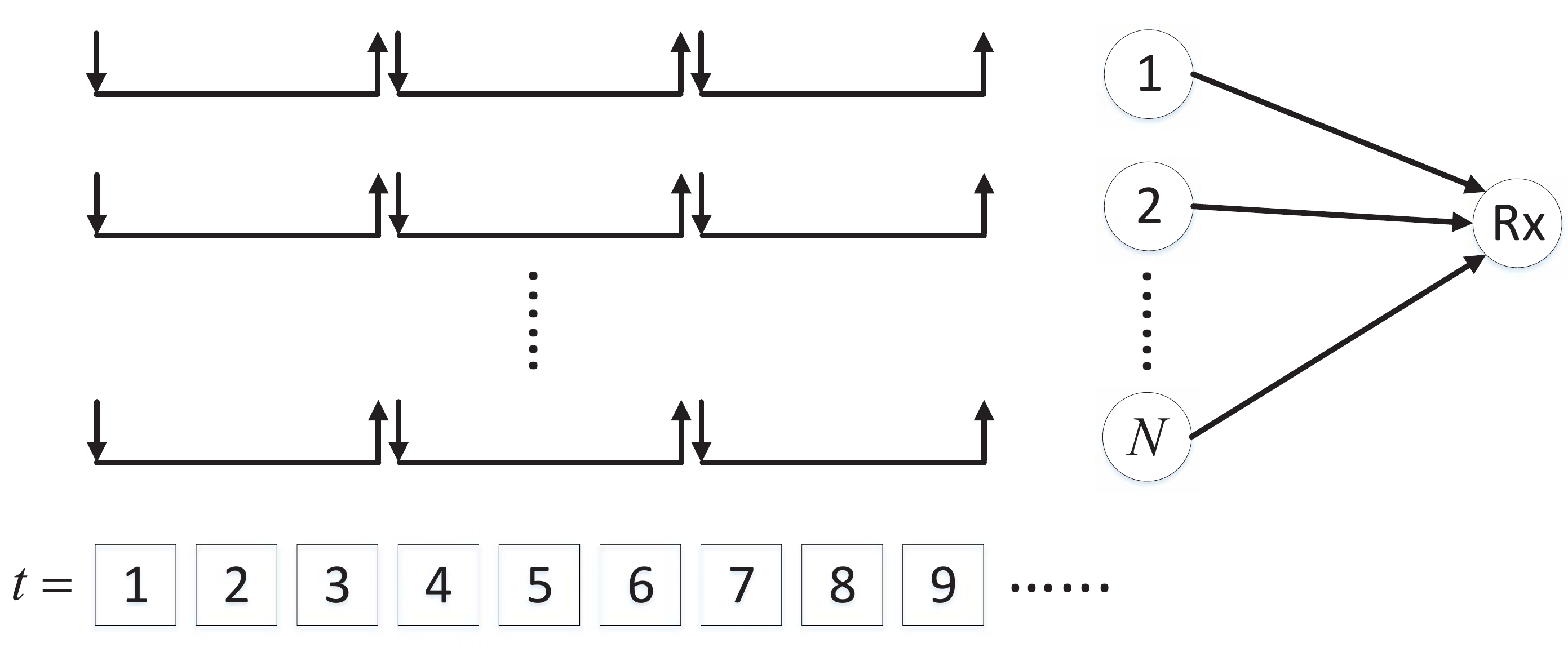}\\
  \caption{System model with $D=3$. }\label{fig:model} \vspace{-0.8cm}
\end{figure}

We consider $N$ stations which share a common wireless channel
and they need to send delay-constrained data packets to a common receiver via the shared wireless channel, as shown in Fig.~\ref{fig:model}.
All packets have the same size and time is slotted where
the slot duration is the time to transmit a packet from a station to the receiver
and receive the acknowledge from the receiver (or observe the channel output)
about whether the packet has been successfully delivered or not.
Such acknowledge model is common in the literature for slotted ALOHA \cite{hajek1982decentralized,naware2005stability}.\footnote{
Later in Section~\ref{sec:two-variants}, we will analyze framed slotted ALOHA, which does not need the acknowledge information.}
A station can only send a packet at the beginning of a slot.
If only one station transmits, the packet will be delivered successfully. Otherwise, if two or more stations transmit
in the same slot, a collision happens and all packets will be lost.

We assume that each packet has a hard delay $D \ge 1$. Once arriving at a station (say at slot $t$), a packet will expire and be removed from the system
if it cannot be delivered in $D$ slots (i.e., before slot $t+D$). In the rest of this paper, we call $D$ the hard \emph{delay} to refer to
a time span, while we call $t+D$ the hard \emph{deadline} to refer to a time instance. Such feature is fundamentally different from the delay-unconstrained scenario.
In the delay-constrained scenario, the (arrival) traffic patten greatly influences
the system performance. As an initial study, in this paper, we assume a \emph{frame-synchronized traffic pattern}.
It can find applications in NCSs and CPSs where the system generates the control packets/messsages
periodically \cite{kim2012cyber}, and it was widely investigated in the packet scheduling policy design in
the delay-constrained wireless communication community \cite{hou2009qos,deng2017timely}.
In the frame-synchronized traffic pattern, starting from slot 1, all stations have a packet arrival every $D$ slots (which is also the hard delay of
all packets). Staring from slot 1, every consecutive $D$ slots is called a \emph{frame}.
The first frame is from slot 1 to slot $D$; the second frame is from slot $D+1$ to slot $2D$; and so on.
Every station has a packet arrival at the beginning of a frame and the packet expires at the end of the frame.
Fig.~\ref{fig:model} shows an example of $D=3$ of the frame-synchronized traffic pattern.

When a new packet arrives at a station operating with the traditional slotted ALOHA protocol,
it is transmitted in the upcoming time slot.
If a collision happens, the  packet is backlogged in the station's queue, and it will be retransmitted (as an old packet) in the next slot with probability $p \in (0,1]$.
Thus, at the beginning of each frame (e.g., slot 1) in our system, collisions will certainly occur when $N>1$
because all stations have a new packet arrival.
To avoid this problem, we modify the traditional slotted ALOHA protocol such that
all stations will always transmit/retransmit its (new or old) packet with probability $p$.
For simplicity, $p$ is called \emph{retransmission probability}, which also represents the transmission probability
for new packets. This ALOHA scheme is called \emph{$p$-constant (delay-constrained) slotted ALOHA}.
We will consider two delay-constrained slotted ALOHA variants in Section~\ref{sec:two-variants}.

Given system parameters --- the packet hard delay $D$, the number of stations $N$,
and the retransmission probability $p$,
we define the \emph{system timely throughput} as the per-slot
average number of delivered packets before expiration, i.e.,
\be
R(D,N,p) \triangleq \lim_{t \to \infty} \frac{ \mathbb{E} \left[  \substack{\text{number of packets delivered before } \\ \text{expiration from slot 1 to slot $t$}} \right]}{t}.
\label{equ:def-R-lim}
\ee
Our system timely throughput only counts those packets that have been delivered before expiration and ignores those packets that
expire and are removed from the system after their deadlines.  Sometimes $R(D,N,p)$ is simply called \emph{system throughput}.

Since our traffic pattern is frame-synchronized and the retransmission probabilities of all stations are the same at all slots, it is easy to
see that the average number of delivered packets before expiration is the same for all frames. Thus, we only need to focus on the first frame and
the system timely throughput becomes
\be
R(D,N,p) =  \frac{ \mathbb{E} \left[  \substack{\text{number of packets delivered before } \\ \text{expiration from slot 1 to slot $D$}} \right]}{D}.
\label{equ:def-R}
\ee
Note that \eqref{equ:def-R} also indicates that the limit in \eqref{equ:def-R-lim} exists, and thus \eqref{equ:def-R-lim} is well-defined.

We define the optimal retransmission probability as,
\be
p^*(D,N) \triangleq \arg \max_{p \in (0,1]} R(D, N, p),
\label{equ:def-p-opt}
\ee
which maximizes the system throughput $R(D, N, p)$.
We also denote the maximum system throughput by
\be
R^*(D,N) \triangleq \max_{p \in (0,1]} R(D, N, p) = R(D, N, p^*(D,N)).
\label{equ:def-R-opt} \nnb
\ee

Let us see a special case of $D=1$.
When $D=1$,  every station has a new packet arrival every slot. Then the system throughput becomes
\be
R(1, N, p) = \binom{N}{1} p (1-p)^{N-1} = Np(1-p)^{N-1}. \nnb
\ee
It is easy to see that the optimal retransmission probability
is $p^*(1,N)  = \frac{1}{N}$, and the maximum system throughput is
$ R^*(1,N) = \left(1- \frac{1}{N}\right)^{N-1}.$
Clearly, we have
\be
\lim_{N \to \infty} R^*(1,N) = \frac{1}{e} = 36.8\%, \label{equ:opt-R-lim-D-1} 
\ee
which is the same as that in the delay-unconstrained slotted ALOHA with saturated traffic \cite[Chapter 5.3.2]{kurose2013computer}
where all $N$ stations  transmit/retransmit their packets with probability $p$
and each station always has a new packet arrival once its packet has been delivered successfully.

Beyond this special case of $D=1$, next we propose an algorithm to compute $R(D,N,p)$ for any hard delay $D\ge1$ in Section~\ref{sec:algorithm}.
For any $D \ge 1$, we further analyze the asymptotic performance of $p$-constant slotted ALOHA when the number of stations $N$ goes to infinity
in Section~\ref{sec:asymptotic-performance}.

\section{An Algorithm to Compute $R(D,N, p)$}  \label{sec:algorithm}
We will show an algorithm to compute $R(D,N,p)$ in this section.
Recall that we only need to focus on the first frame from slot 1 to slot $D$ (see \eqref{equ:def-R}).
At (the beginning of) slot 1, all $N$ stations have a new packet arrival.
Random variable $M_t$ denotes  the number of stations that have already delivered their packets (we call them \emph{finished stations}) before slot $t \in \{1,2,\cdots,D\}$.
Clearly $M_1=0$ and $M_t \in \{0,1,2,\cdots,\min\{t-1,N\}\}$.
We also let $M_{D+1}$ denote the number of finished stations at the end of slot $D$.
Note that $N-M_t$ is the number of \emph{active stations} at (the beginning of) slot $t$  that will transmit/retransmit
their packets with probability $p$ in slot $t$.
Random variable $S_t\in \{0,1\}$ denotes the number of packets delivered successfully in slot $t \in \{1,2,\cdots,D\}$.
Thus, the total number of packets delivered successfully in the first frame
is $S_1 + S_2+\cdots + S_D$. According to \eqref{equ:def-R}, the system throughput is
\bee
& R(D,N,p) = \frac{\mathbb{E}[S_1 + S_2 + \cdots + S_D]}{D}   = \frac{\mathbb{E}[S_1] + \mathbb{E}[S_2] + \cdots + \mathbb{E}[S_D] }{D}
= \frac{\sum_{t=1}^D P(S_t=1) }{D}.
\label{equ:algo-R-based-on-S-t}
\eee
To calculate $R(D,N,p)$, we only need to calculate $P(S_t=1), \forall t \in \{1,2,\cdots,D\}$.
Note that
\be
P(S_t = 1) = \sum_{m=0}^{\min\{t-1,N\}} P(S_t=1|M_t=m) P(M_t=m),
\label{equ:algo-prob-S-t=1}
\ee
and
\bee
& P(S_t=1|M_t=m)
= \left\{
  \begin{array}{ll}
    (N-m)p(1-p)^{N-m-1}, & \hbox{if $m<N$;} \\
    0, & \hbox{if $m=N$.}
  \end{array}
\right.
\label{equ:algo-cond-prob-S-t-M-t}
\eee
Thus, to obtain $P(S_t = 1)$, we only need to calculate
\bee
& P(M_t=m), \quad \forall t \in \{1,2,\cdots,D\},  \forall m \in \{0,1,2,\cdots,\min\{t-1,N\}\}. \label{equ:algo-P-M-t-m-all-t-all-m}
\eee
The evolution of $M_t$ can be described as a trellis,
as illustrated in Fig.~\ref{fig:trellis}
for $N=2$ and $D=4$.
Based on the trellis, we can recursively calculate $P(M_t=m)$ as follows:
\begin{itemize}
\item $t=1$: $P(M_t=0)=P(M_1=0)=1.$
\item For any $t \in \{1,2,\cdots, D-1\}$, based on $P(M_t=m)$, we calculate $P(M_{t+1}=m)$ where $m \in \{0,1,\cdots, \min\{t,N\}\}$
according to the following three cases:

\begin{figure}
  \centering
  \includegraphics[width=0.6\linewidth]{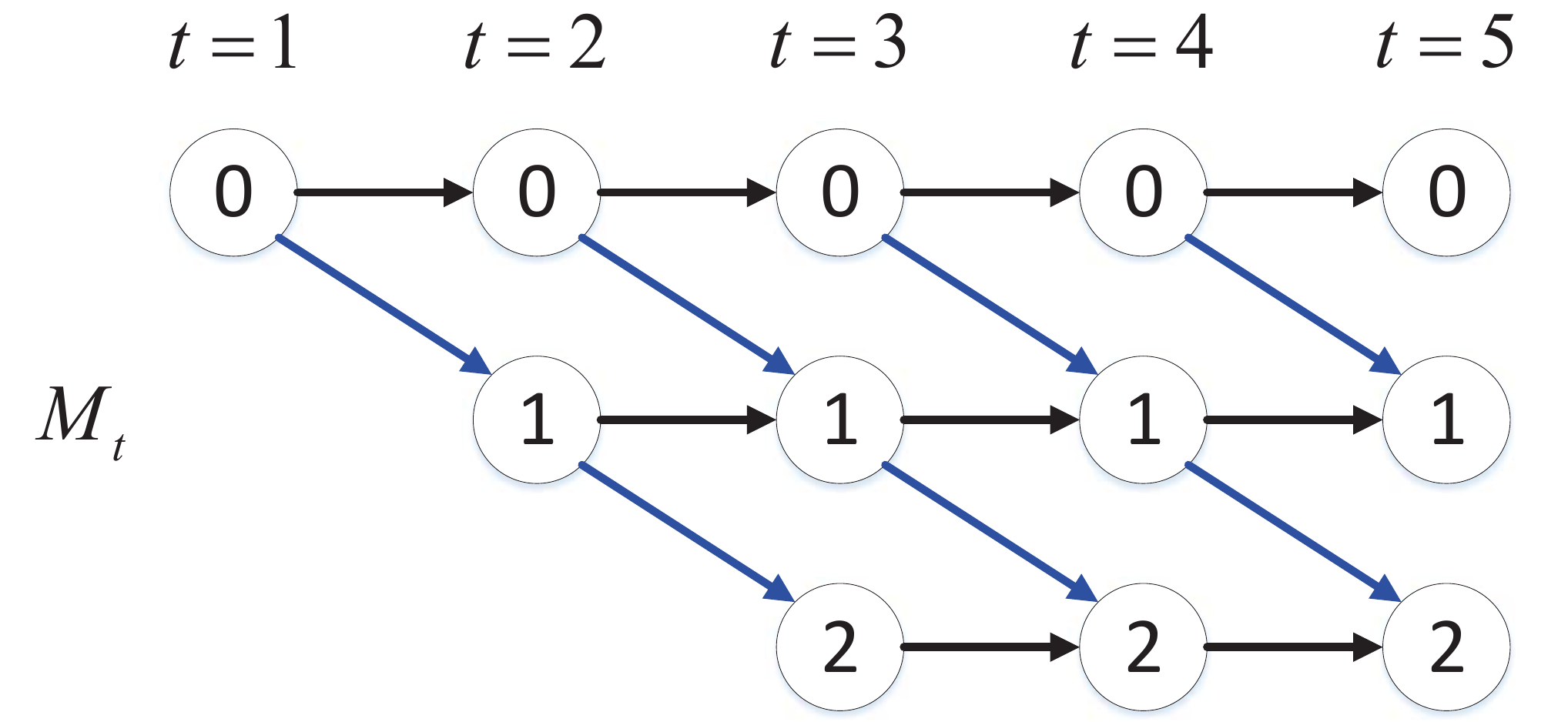}\\
  \caption{The trellis describing the evolution of $M_t$ when $N=2$ and $D=4$. Note that
we also show $M_{D+1}=M_5$, which is the number of finished stations at the end of slot $D$.
Any blue link in the trellis corresponds to a successful delivery, implying $S_t=1$.
}\label{fig:trellis} \vspace{-0.8cm}
\end{figure}

(i) When $m=0$, we have
\bee
& P(M_{t+1}=m)=P(M_{t+1}=0) = \sum_{m'=0}^{\min\{t-1,N\}} P(M_{t+1}=0|M_{t}=m') P(M_{t}=m') \nnb \\
& = P(M_{t+1}=0|M_{t}=0) P(M_{t}=0)   = \left[1-Np(1-p)^{N-1}\right]P(M_{t}=0).
\label{equ:alg-t+1-m=0}
\eee

(ii) When $m \in \{1,2,\cdots, \min\{t, N\}-1\}$, we have
\bee
& P(M_{t+1}=m)  = \sum_{m'=0}^{\min\{t-1,N\}} P(M_{t+1}=m|M_{t}=m') P(M_{t}=m') \nnb \\
& = P(M_{t+1}=m|M_{t}=m-1) P(M_{t}=m-1)   + P(M_{t+1}=m|M_{t}=m) P(M_{t}=m)\nnb \\
& \resizebox{0.92\linewidth}{!}{$=  \left[ N-(m-1) \right]p(1-p)^{N-(m-1)-1} P(M_{t}=m-1) + \left[1-(N-m)\right]p(1-p)^{N-m-1}P(M_{t}=m)$}. \label{equ:alg-t+1-m=1-to-t-1}
\eee

(iii) When $m =\min\{t, N\}$, we consider two sub-cases:
\begin{itemize}
\item When $t \le N$, we have $m =\min\{t, N\}=t$ and
\bee
& P(M_{t+1}=m) = P(M_{t+1}=t)   = \sum_{m'=0}^{\min\{t-1,N\}} P(M_{t+1}=t|M_{t}=m') P(M_{t}=m') \nnb \\
& = P(M_{t+1}=t|M_{t}=t-1) P(M_{t}=t-1)  \nnb \\
& =  \left[N-(t-1)\right]p(1-p)^{N-(t-1)-1}  P(M_{t}=t-1). \label{equ:alg-t+1-m=t-subcase1}
\eee
\item When $t>N$, we have $m =\min\{t, N\}=N$ and
\bee
& P(M_{t+1}=m) = P(M_{t+1}=N)  = \sum_{m'=0}^{\min\{t-1,N\}} P(M_{t+1}=N|M_{t}=m') P(M_{t}=m') \nnb \\
& = P(M_{t+1}=N|M_{t}=N-1) P(M_{t}=N-1)  + P(M_t=N) \nnb \\
& = p \cdot P(M_{t}=N-1) + P(M_t=N).
\label{equ:alg-t+1-m=t-subcase2}
\eee
\end{itemize}

\end{itemize}

Therefore, we can first use \eqref{equ:alg-t+1-m=0}-\eqref{equ:alg-t+1-m=t-subcase2} to recursively obtain \eqref{equ:algo-P-M-t-m-all-t-all-m},
whose time complexity is $O( \min\{D,N\} \cdot D)$. Then we use \eqref{equ:algo-cond-prob-S-t-M-t} and \eqref{equ:algo-P-M-t-m-all-t-all-m}
to obtain $P(S_t=1)$, i.e., \eqref{equ:algo-prob-S-t=1}, for all $t=1,2,\cdots, D$, whose time complexity is $O( \min\{D,N\} \cdot D)$.
Finally, we use \eqref{equ:algo-R-based-on-S-t} to compute the system throughput $R(D,N,p)$, whose time complexity is $O(D)$.
The overall time complexity to compute the system throughput $R(D,N,p)$ is $O( \min\{D,N\} \cdot D)$.

Now we can use this algorithm to compute $R(D,N,p)$ for any system parameters $D,N$ and $p$,
and thus we can numerically  obtain the optimal retransmission probability $p^*(D,N)$ and the maximum system
throughput $R^*(D,N)$ (with certain step-size error).

\section{Asymptotic Performance of $p$-Constant ALOHA} \label{sec:asymptotic-performance}
For $p$-constant ALOHA, we have shown that the maximum system throughput converges to $1/e$ when $D=1$ in \eqref{equ:opt-R-lim-D-1}.
This result is the same as that in delay-unconstrained slotted ALOHA with saturated traffic.
For general (fixed) $D\ge 1$, when $N$ is large, at slot 1, it has $N$ active stations that have packets to send;
at slot 2, there are at least $N-1$ active stations.
Similarly, at any slot $t \in \{1,2,\cdots,D\}$, there are at least $N-t+1$ active stations. Since $N-t+1\rightarrow \infty$ (as $N\rightarrow \infty$) for $1\leq t\leq D$, it is reasonable to anticipate that
a similar situation to the delay-unconstrained case with saturated traffic
occurs in the delay-constrained case --- the number of active stations  goes unbounded
at all slots.
According to this intuition, we conjecture that the maximum system throughput converges to $1/e=36.8\%$
for any hard delay $D$. We now present one of our main results.

\begin{theorem} \label{thm:converge-to-1-over-e}
For any $D \ge 1$,  we have $\lim_{N \to \infty} R^*(D,N) = 1/e.$
\end{theorem}
Next we prove Theorem~\ref{thm:converge-to-1-over-e} rigourously.
Towards that end, we first prove some useful lemmas.

\subsection{Useful Lemmas}
The optimal retransmission probability $p^*(D, N)$ depends on the number of stations $N$ and the
packet hard delay $D$. In this section, we fix $D$ but evaluate the asymptotic performance when $N \to \infty$.
For notational convenience,
we drop the dependence of the retransmission probability on $D$
and let $p(N) \in (0,1]$ denote  the retransmission probability when the number of stations is $N$.
In the paper, we assume that all stations know $N$, and are able to choose the retransmission probability
depending on $N$.
In practice, when $N$ is not known a priori,
there are many works to estimate $N$ based on the acknowledge information,
e.g., \cite{stefanovic2013joint, zhang2017optimal},
which can be applied to our problem.  Later, we will also propose an algorithm to estimate $N$ in Algorithm~\ref{alg:prediction}.
For the time being, we assume that $N$ is known a priori.

\begin{lemma} \label{lem:a-key-lemma}
Consider a sequence $\{N_k\}$ of positive integers where $N_1 < N_2 < N_3 < \cdots$, which
is a subsequence of the positive-integer sequence $\{1,2,3,\cdots\}$.
If the retransmission probability $p(N)$ satisfies
\be
\lim_{k \to \infty} p(N_k) = 0, \label{equ:p(N)-to-0}
\ee
and
\be
\lim_{k \to \infty} N_k \cdot p(N_k) = \delta, \label{equ:Np(N)-to-1}
\ee
where $\delta$ could be any non-negative real number or $\infty$,
then
\be
\lim_{k \to \infty}  (N_k-A) p(N_k) \left[1- p(N_k)\right]^{N_k-B} = \frac{\delta}{e^{\delta}},
\label{equ:limit-is-delta-over-exp-delta}
\ee
where $A$ and $B$ are  any non-negative integers and by convention we set
\be
\frac{\infty}{e^{\infty}} = 0.
\label{equ:convention}
\ee
\end{lemma}
\begin{IEEEproof}
\ifx \ISTR \undefined
Please see Appendix~\ref{app:proof-of-lem-a-key-lemma} in our technical report \cite{TR}. 
\else
Please see Appendix~\ref{app:proof-of-lem-a-key-lemma}.
\fi
\end{IEEEproof}

Now we consider the first frame from slot 1 to slot $D$. Similar to the definition of random variable $M_t$ in Section~\ref{sec:algorithm},
we let random variable $M_{N,t}$ denote the number of finished stations, i.e., those stations that have already delivered their packets,  before slot $t \in \{1,2,\cdots,D\}$
when the total number of stations is $N$.
The reason that we introduce a new notation $M_{N,t}$ is because we need to explore its relation with respect to $N$.
Clearly $M_{N,t} \in \{0,1,2,\cdots, \min\{t-1,N\}\}$. Later when we prove the sequence-limit-related statements, we
consider sufficiently large $N$, e.g., $N \ge D+1$. And thus $\min\{t-1,N\}=t-1$.
Then
by denoting the probability of $M_{N,t}=m$ by $P(M_{N,t}=m)$,
we prove the following lemma.
\begin{lemma} \label{lem:prob-m-t-limit-exist}
Consider a sequence $\{N_k\}$ of positive integers where $N_1 < N_2 < N_3 < \cdots$, which
is a subsequence of the positive-integer sequence $\{1,2,3,\cdots\}$.
If the retransmission probability $p(N_k)$ satisfies \eqref{equ:p(N)-to-0} and \eqref{equ:Np(N)-to-1}
where $\delta$ could be any non-negative real number or $\infty$,
then the sequence $\{P(M_{N_k,t} = m)\}_{k=1}^{\infty}$ has a limit for any $t \in \{1,2,\cdots,D\}$ and any $m \in \{0,1,2,\cdots, t-1\}$.
Namely, there exists a non-negative real number $\gamma_t^m$ such that\footnote{Note that $\gamma_t^m$ is a notation with both subscript $t$ and superscript $m$. It should not
be understood as $(\gamma_t)^m$.}
\bee
& \lim_{k \to \infty} P(M_{N_k,t} = m) = \gamma_t^m,  \forall t \in \{1,2,\cdots,D\}, \forall m \in \{0,1,\cdots, t-1\}. \label{equ:P-M-N-t=m-limit-exist}
\eee
In addition,
\be
\sum_{m=0}^{t-1} \gamma_t^m = 1, \quad \forall t \in \{1,2,\cdots,D\}. \label{equ:limit-sum-is-1}
\ee
\end{lemma}
\begin{IEEEproof}
\ifx \ISTR \undefined
Please see Appendix~\ref{app:proof-of-prob-m-t-limit-exist} in our technical report \cite{TR}.
\else
Please see Appendix~\ref{app:proof-of-prob-m-t-limit-exist}.
\fi
\end{IEEEproof}

We further present another preliminary lemma.

\begin{lemma} \label{lem:an-bn-limsup}
Suppose that $\{a_N\}, \{b_N\}$ are two bounded sequences where $b_N \ge 0, \forall N$. If
$\limsup_{N \to \infty} a_N \le 0,$
then
$
\limsup_{N \to \infty} a_N b_N \le 0.
$
\end{lemma}
\begin{IEEEproof}
\ifx \ISTR \undefined
Please see Appendix~\ref{app:proof-of-lem-an-bn-limsup} in our technical report \cite{TR}.
\else
Please see Appendix~\ref{app:proof-of-lem-an-bn-limsup}.
\fi
\end{IEEEproof}

\subsection{The System Throughput when $p(N)=1/N$}
In this subsection, we leverage Lemma~\ref{lem:prob-m-t-limit-exist} to
show that the system throughput goes to $1/e$ when the retransmission probability is $p(N)=1/N$.

\begin{lemma} \label{lem:p-1-N-to-1-e}
Consider the retransmission probability $p(N) = 1/N$ when the number of stations is $N$.
Then the system throughput converges to $1/e$, i.e.,
\be
\lim_{N \to \infty} R(D, N, 1/N) = 1/e. \nnb
\ee
\end{lemma}
\begin{IEEEproof}
\ifx \ISTR \undefined
Please see Appendix~\ref{app:proof-of-lem-p-1-N-to-1-e} in our technical report \cite{TR}.
\else
Please see Appendix~\ref{app:proof-of-lem-p-1-N-to-1-e}.
\fi
\end{IEEEproof}

\subsection{Proof of Theorem~\ref{thm:converge-to-1-over-e}}
We then proceed to prove Theorem~\ref{thm:converge-to-1-over-e} when we use the optimal retransmission probability $p(N) = p^*(D,N)$.
Since $p^*(D,N)$ is the optimal retransmission probability to maximize the system throughput, we have
\be
R^*(D,N) = R(D,N, p^*(D,N)) \ge R(D, N, 1/N). \nnb
\ee
Thus, according to Lemma~\ref{lem:p-1-N-to-1-e}, we have
\be
\liminf_{N \to \infty} R^*(D,N) \ge \liminf_{N \to \infty} R(D, N, 1/N) = 1/e. \nnb
\ee
Then Theorem~\ref{thm:converge-to-1-over-e} holds if we can
verify
\be
\limsup_{N \to \infty} R^*(D,N)  \le 1/e. \label{equ:thm-1-over-e-proof-equ1}.
\ee

For any slot $t \in \{1,2,\cdots,D\}$ in the first frame,
suppose that we have $M_{N,t}=m  \in \{0, 1, \cdots, t-1\}$ finished stations  before slot $t$.
Clearly, there are $N-m$ active stations that have a packet
at slot $t$. Then the probability of delivering a packet in slot $t$ given that there are $m$ finished station before slot $t$ is
\bee
\resizebox{\linewidth}{!}{$ \alpha(N,m)  = (N-m)p^*(N,D) \left[1-p^*(N,D) \right]^{N-m-1}  \le \max_{p \in (0,1]} \left[  (N-m) p (1-p)^{N-m-1} \right]
 =  \left(1- \frac{1}{N-m} \right)^{N-m-1}$}, \nnb
\eee
where the last equality follows from the fact that
\be
\arg \max_{p \in (0, 1]} p (1-p)^{N-m-1}   = \frac{1}{N-m}. \nnb
\ee
Then,  we have
\be
\limsup_{N \to \infty} \alpha(N,m) \le \limsup_{N \to \infty}  \left(1- \frac{1}{N-m} \right)^{N-m-1}  = \frac{1}{e}, \nnb
\ee
implying that
\be
\limsup_{N \to \infty} \left[ \alpha(N,m) - \frac{1}{e} \right] \le 0. \nnb
\ee

Recall that we use random variable $S_{N,t} \in \{0,1\}$ to denote the number of packets delivered in slot $t$.
Then, the probability of delivering a packet in slot $t$ is $P(S_{N,t}=1)$, and we have
\bee
& P(S_{N,t}=1) - \frac{1}{e}  = \sum_{m=0}^{t-1} P(S_{N,t}=1|M_{N,t}=m) P(M_{N,t}=m)  - \frac{1}{e}\nnb \\
& = \sum_{m=0}^{t-1} \alpha(N,m) P(M_{N,t}=m) - \frac{1}{e} \cdot \sum_{m=0}^{t-1} P(M_{N,t}=m) = \sum_{m=0}^{t-1} \left[\alpha(N,m) - \frac{1}{e} \right] P(M_{N,t}=m). \nnb
\eee
If we set $a_N = \alpha(N,m) - \frac{1}{e}, b_N = P(M_{N,t}=m)$, we can see that
$\{a_N\}$ and $\{b_N\}$ satisfy the conditions in Lemma~\ref{lem:an-bn-limsup}. Hence,
\be
\limsup_{N \to \infty} a_N b_N = \limsup_{N \to \infty} \left[\alpha(N,m) - \frac{1}{e} \right] P(M_{N,t}=m) \le 0. \nnb
\ee
Then, we have
\bee
& \limsup_{N \to \infty} P(S_{N,t}=1) - \frac{1}{e}   = \limsup_{N \to \infty} \left[P(S_{N,t}=1) - \frac{1}{e} \right]  = \limsup_{N \to \infty}  \sum_{m=0}^{t-1} \left[\alpha(N,m) - \frac{1}{e} \right] P(M_{N,t}=m) \nnb \\
& \le  \sum_{m=0}^{t-1} \left\{\limsup_{N \to \infty}   \left[\alpha(N,m) - \frac{1}{e} \right] P(M_{N,t}=m) \right\} \le  \sum_{m=0}^{t-1} 0  = 0. \nnb
\eee
Therefore, we have
\be
\limsup_{N \to \infty} P(S_{N,t}=1) \le \frac{1}{e}. \nnb
\ee
Thus, the system throughput $R^*(D,N)$ satisfies
\bee
& \limsup_{N \to \infty} R^*(D,N) = \limsup_{N \to \infty} \frac{\sum\limits_{t=1}^{D} P(S_{N,t} = 1)}{D}  \le \frac{\sum\limits_{t=1}^{D} \limsup\limits_{N \to \infty}  P(S_{N,t} = 1) }{D}  \le \frac{\sum\limits_{t=1}^{D} \frac{1}{e}}{D} = \frac{1}{e}, \nnb
\eee
which proves \eqref{equ:thm-1-over-e-proof-equ1}.
The proof for Theorem~\ref{thm:converge-to-1-over-e} is completed.

\subsection{The Asymptotic Behavior of the Optimal Retransmission Probability}
Theorem~\ref{thm:converge-to-1-over-e} shows that the asymptotic maximum system throughput converges to $1/e=36.8\%$ for
any hard delay $D$. This is one key indicator for asymptotic performance that we hope to understand for delay-constrained slotted ALOHA.
It suggests that the asymptotic maximum system throughput for delay-constrained case is the same as that for delay-unconstrained case with saturated traffic.
Another indicator for asymptotic performance is how the optimal retransmission probability $p^*(D,N)$ changes as $N$ goes to infinity.
Our Lemma~\ref{lem:p-1-N-to-1-e} shows that when $p(N)=1/N$, the system throughput converges to $1/e$, which is exactly the asymptotic maximum
system throughput (achieved by the optimal retransmission probability $p^*(D,N)$). It is thus reasonable to conjecture that
$p^*(D,N)$ behaves as $1/N$ when $N$ is large enough. To prove our conjecture (Theorem~\ref{thm:opt-p-1-N}), we first prove the following two lemmas.
\begin{lemma} \label{lem:n-p-not-1-lead-to-suboptimal-rate}
Consider a sequence $\{N_k\}$ of positive integers where $N_1 < N_2 < N_3 < \cdots$, which
is a subsequence of the positive-integer sequence $\{1,2,3,\cdots\}$.
If the retransmission probability $p(N)$
satisfies \eqref{equ:p(N)-to-0} and \eqref{equ:Np(N)-to-1}
where $\delta$ could be any non-negative real number or $\infty$ but $\delta \neq 1$,
then the limit $\lim_{k \to \infty} R(D, N_k, p(N_k))$ exists and satisfies
\be
\lim_{k \to \infty} R(D, N_k, p(N_k))  < \frac{1}{e}. \nnb
\ee
\end{lemma}
\begin{IEEEproof}
\ifx \ISTR \undefined
Please see Appendix~\ref{app:proof-of-lem-n-p-not-1-lead-to-suboptimal-rate} in our technical report \cite{TR}.
\else
Please see Appendix~\ref{app:proof-of-lem-n-p-not-1-lead-to-suboptimal-rate}.
\fi
\end{IEEEproof}

Now we provide another lemma to show that the optimal retransmission probability converges to 0 as the number of stations goes to infinity.
\begin{lemma} \label{lem:opt-p-converge-to-0}
For any $D \ge 1$, we have $\lim\limits_{N \to \infty} p^*(D,N) = 0.$
\end{lemma}
\begin{IEEEproof}
\ifx \ISTR \undefined
Please see Appendix~\ref{app:proof-of-lem-opt-p-converge-to-0} in our technical report \cite{TR}.
\else
Please see Appendix~\ref{app:proof-of-lem-opt-p-converge-to-0}.
\fi
\end{IEEEproof}

We then prove the asymptotic performance of the optimal retransmission probability.
\begin{theorem} \label{thm:opt-p-1-N}
For any $D \ge 1$, we have
$
\lim\limits_{N \to \infty} N\cdot p^*(D,N)=1. $
\end{theorem}
\begin{IEEEproof}
We use  contradiction to prove the following two results:
\be
\limsup_{N \to \infty} Np^*(D, N) \le 1,
\label{equ:limsup-n-opt-p-less-than-1}
\ee
and
\be
\liminf_{N \to \infty} Np^*(D, N) \ge 1.
\label{equ:liminf-n-opt-p-larger-than-1}
\ee
If both \eqref{equ:limsup-n-opt-p-less-than-1} and \eqref{equ:liminf-n-opt-p-larger-than-1}
hold, we have that $\lim_{N \to \infty} Np^*(D, N) = 1$, and thus finish the proof.

\textbf{Proof of \eqref{equ:limsup-n-opt-p-less-than-1}}.
Suppose $\limsup_{N \to \infty} Np^*(D, N) > 1$. Then we can find a subsequence $\{N_k\}_{k=1}^{\infty}$
such that
\be
\lim_{k \to \infty} N_k p^*(D, N_k) = \delta_1 > 1,
\label{equ:NpN-delta-1}
\ee
where $\delta_1$ could be $\infty$.
Lemma~\ref{lem:opt-p-converge-to-0} shows that $p^*(D,N_k)$ satisfies \eqref{equ:p(N)-to-0}.
In addition, \eqref{equ:NpN-delta-1} shows that $p^*(D,N_k)$ satisfies \eqref{equ:Np(N)-to-1} with $\delta=\delta_1 > 1$.
Then Lemma~\ref{lem:n-p-not-1-lead-to-suboptimal-rate} shows that
\be
\lim_{k \to \infty} R^*(D, N_k)=\lim_{k \to \infty} R(D, N_k, p^*(D, N_k))  < \frac{1}{e}, \nnb
\ee
which contradicts to Theorem~\ref{thm:converge-to-1-over-e}. Thus, \eqref{equ:limsup-n-opt-p-less-than-1} holds.

\textbf{Proof of \eqref{equ:liminf-n-opt-p-larger-than-1}.}
Suppose $\liminf_{N \to \infty} Np^*(D, N) < 1$. Then we can find a subsequence $\{N_k\}_{k=1}^{\infty}$
such that
\be
\lim_{k \to \infty} N_k p^*(D, N_k) = \delta_2 < 1. \nnb
\ee
Again, from Lemma ~\ref{lem:n-p-not-1-lead-to-suboptimal-rate}, we have
\be
\lim_{k \to \infty} R^*(D, N_k) = \lim_{k \to \infty} R(D, N_k, p^*(D, N_k))  < \frac{1}{e}, \nnb
\ee
which contradicts to Theorem~\ref{thm:converge-to-1-over-e}. Thus, \eqref{equ:liminf-n-opt-p-larger-than-1} also holds.

The proof is completed.
\end{IEEEproof}

Theorem~\ref{thm:opt-p-1-N} rigorously shows that indeed the optimal retransmission probability behaves asymptotically as $1/N$.
Thus, when $N$ is large enough, even though we cannot obtain the  explicit formula for the optimal retransmission probability,
we can simply let it be $1/N$, which  simplifies the system design for $p$-constant delay-constrained slotted ALOHA.

\section{Two Delay-Constrained Slotted ALOHA Variants} \label{sec:two-variants}
Previous sections considered $p$-constant (delay-constrained) slotted ALOHA where the retransmission probability
is the same all the time. In this section, we analyze two delay-constrained slotted ALOHA variants: $p$-dynamic
slotted ALOHA and framed slotted ALOHA.
\subsection{$p$-dynamic Slotted ALOHA}
Instead of only considering constant retransmission probability,
there is another design space of dynamically changing the retransmission probability at different slots,
which is called \emph{$p$-dynamic slotted ALOHA}.
We assume that each station knows the number of active stations (i.e., those stations with a not-yet-delivered packet) at
 slot $t$, which is denoted by $n(t)$.

Let $p_{\textsf{dynamic}}(n,t)$ denote the retransmission probability in slot $t$ when there are $n \ge 1$ active stations\footnote{If there are no active stations, all stations will remain idle
and hence $p_{\textsf{dynamic}}(0,t)=0$. We thus ignore this degenerate case in the rest of this subsection.} at the beginning of slot $t$.
Due to the frame-synchronized structure, we only need to consider the first frame, i.e., $t=1,2,\cdots,D$.
We then vectorize $p_{\textsf{dynamic}}(n_t,t)$ as $\bm{p}= \{p_{\textsf{dynamic}}(n,t), n=1,2,\cdots,N, t=1,2,\cdots, D \}$.
Note that for a given retransmission  policy $\bm{p}$, we can use the similar algorithm in Section~\ref{sec:algorithm}
to compute the system throughput, denoted by $R_{\textsf{dynamic}}(D, N, \bm{p})$.
In addition, we are interested in finding the best retransmission policy $\bm{p}$ to maximize the system
throughput, i.e.,
\be
\bm{p}^* = \arg \max_{\bm{p}} R_{\textsf{dynamic}}(D,N, \bm{p}). \nnb
\ee

Similar to the analysis in Section~\ref{sec:algorithm}, we let  $S_t \in \{0,1\}$ denote the number of packets delivered
successfully in slot $t \in \{1,2,\cdots,D\}$.
Then similar to \eqref{equ:algo-R-based-on-S-t}, we can compute the system throughput, i.e.,
\bee
R_{\textsf{dynamic}}(D,N, \bm{p}) =  \frac{\sum_{t=1}^D P(S_t=1)}{D}, \nnb
\eee
where the distribution of $S_t$ depends on system parameters $D$ and $N$ and the retransmission policy $\bm{p}$.
Then similar to \eqref{equ:algo-prob-S-t=1} and \eqref{equ:algo-cond-prob-S-t-M-t}, we have
\be
P(S_t=1) = \sum_{n=1}^N P(S_t=1|n(t)=n) P(n(t)=n), \nnb
\ee
and
\bee
& P(S_t=1|n(t)=n)  = n \cdot p_{\textsf{dynamic}}(n,t) \cdot \left[ 1 - p_{\textsf{dynamic}}(n,t) \right]^{n-1} \le \left( 1 - \frac{1}{n}\right)^{n-1}, \nnb
\eee
where the inequality is achieved when
\be
p_{\textsf{dynamic}}(n,t) = \frac{1}{n}. \label{equ:opt-p-dynamic-1}
\ee
Thus, among all possible retransmission policies, the one in \eqref{equ:opt-p-dynamic-1}
maximizes $P(S_t=1)$, and then maximizes the system throughput $R_{\textsf{dynamic}}(D,N, \bm{p})$.
Therefore, the optimal retransmission policy $\bm{p}^*$ is
\be
p^*_{\textsf{dynamic}}(n,t) = \frac{1}{n}.
\label{equ:opt-p-dynamic-2} \nnb
\ee
Note that this retransmission policy is stationary for all slots. Then as we have used  $n(t)$ to denote
the number of active stations at slot $t$, we can rewrite the optimal retransmission policy as
\be
p^*_{\textsf{dynamic}}(n(t)) = \frac{1}{n(t)}.
\label{equ:opt-p-dynamic-3}
\ee

We further let  $R^*_{\textsf{dynamic}}(D,N) = R_{\textsf{dynamic}}(D,N, \bm{p}^*)$ denote the maximum
system throughput of $p$-dynamic slotted ALOHA. We show its asymptotic system throughput.
\begin{theorem} \label{thm:opt-R-converge-to-1-over-e-dynamic}
For any $D \ge 1$,
\be
\lim_{N \to \infty} R^*_{\textsf{dynamic}}(D, N) = \lim_{N \to \infty} R_{\textsf{dynamic}}(D, N, \bm{p}^*)= 1/e.
\ee
\end{theorem}
\begin{IEEEproof}
\ifx \ISTR \undefined
Please see Appendix~\ref{app:proof-of-thm-opt-R-converge-to-1-over-e-dynamic} in our technical report \cite{TR}.
\else
Please see Appendix~\ref{app:proof-of-thm-opt-R-converge-to-1-over-e-dynamic}.
\fi
\end{IEEEproof}

Theorem~\ref{thm:opt-R-converge-to-1-over-e-dynamic} shows that the design space of dynamic retransmission probability
cannot enlarge the asymptotic system throughput. Thus, to achieve the best asymptotic system throughput, i.e., $1/e$,
it suffices to use the constant retransmission probability.
Although $p$-constant slotted ALOHA and $p$-dynamic slotted ALOHA
have the same asymptotic system throughput, they are different in two aspects.
First, $p$-constant slotted ALOHA only needs to know the total number of stations, i.e., $N$, to compute
the optimal retransmission probability $p^*(D,N)$ while $p$-dynamic slotted ALOHA needs to know the number
of active stations at each slot, i.e, $n(t)$, to compute the optimal retransmission policy, i.e., \eqref{equ:opt-p-dynamic-3}.
Second, $p$-constant slotted ALOHA  needs to run an algorithm to compute the optimal retransmission probability $p^*(D,N)$
but the optimal retransmission policy of $p$-dynamic slotted ALOHA has a closed-form, i.e., \eqref{equ:opt-p-dynamic-3},
which  simplifies the computation.

\subsection{Framed Slotted ALOHA} \label{subsec:framed}
We further consider another different slotted ALOHA scheme for the frame-synchronized traffic pattern,
which is called \emph{framed slotted ALOHA} \cite{lee2005enhanced,prodanoff2010optimal,yu2017stability}.
At the beginning of each frame, say the first frame,
each station randomly picks up a slot, say slot $t$, according to a uniform distribution; then the station transmits
its packet with probability $p$ at slot $t$, and remains idle at all other slots in this frame.
Note that since each station will at most transmit once in a frame, this scheme does not
require the acknowledge information of the receiver/channel, and all stations will not retransmit their packets. Random variable $S_t$ denotes  the number of packets delivered
in slot $t \in \{1,2,\cdots,D\}$.
Then the system throughput is
\bee
& R_{\textsf{framed}} (D,N, p) =  \frac{\mathbb{E}[S_1 + S_2 + \cdots S_D]}{D}  = \frac{\sum_{t=1}^D P(S_t=1)}{D} = P(S_1=1), \nnb
\eee
where the last equality holds because all slots will be selected by any station with equal probability, and there is
no retransmission in the same frame. We let $p^* = \arg\max_{p \in (0,1]} R_{\textsf{framed}} (D,N, p)$ denote
the optimal retransmission probability to maximize the system throughput of framed slotted ALOHA, and
let $R^*_{\textsf{framed}} (D,N)=R_{\textsf{framed}} (D,N, p^*)$ denote the corresponding maximum system throughput.

Note that when $D=1$, we have
\bee
P(S_t=1) &= Np(1-p)^{N-1}, \nnb
\eee
which is maximized at $p=1/N$.
 When $D\ge 2$, we let $N_t$ denote  the number of stations picking up slot $t \in \{1,2,\cdots,D\}$ and thus we have
\bee
& P(S_t=1)  =  \sum_{k=0}^N P(S_t=1|N_t=k)P(N_t=k)
 = \sum_{k=0}^N kp(1-p)^{k-1} \binom{N}{k} \left(\frac{1}{D} \right)^k \left(1 - \frac{1}{D} \right)^{N-k}  \nnb \\
& = \resizebox{\linewidth}{!}{$\frac{p}{1-p} \cdot \left(\frac{D-p}{D} \right)^N \sum_{k=0}^{N} k \binom{N}{k} \left( \frac{1-p}{D-p} \right)^k \left( \frac{D-1}{D-p} \right)^{N-k}
 = \frac{p}{1-p} \cdot \left(\frac{D-p}{D} \right)^N  \cdot N \cdot \frac{1-p}{D-p}  = \frac{pN}{D-p} \cdot \left(\frac{D-p}{D} \right)^N.$} \nnb
\eee
Taking derivative with respect to $p$ yields
\[
\frac{ND}{(D-p)^2} \left(\frac{D-p}{D} \right)^N  + \frac{pN}{D-p} \cdot N \left(\frac{D-p}{D} \right)^{N-1}
\cdot \left(-\frac{1}{D}\right) = 0,
\]
which leads to
\be
p^* =
\left\{
  \begin{array}{ll}
    \frac{D}{N}, & \hbox{if $D<N$;} \\
    1, & \hbox{otherwise.}
  \end{array}
\right.
= \min\left\{ \frac{D}{N}, 1 \right\}.
\label{equ:opt-p-framed}
\ee
Note that \eqref{equ:opt-p-framed} also includes $p^*=1/N$ for the case of $D=1$. Thus,
\eqref{equ:opt-p-framed} is the optimal retransmission probability of framed slotted ALOHA for any given $D$ and $N$.

Thus, when $N > D > 1$, $p^*=D/N$ and the maximum system throughput is
\bee
\resizebox{0.88\linewidth}{!}{$ R^*_{\textsf{framed}} (D,N)=R_{\textsf{framed}} (D,N, p^*) = P(S_t=1) = \frac{p^*N}{D-p^*} \cdot \left(\frac{D-p^*}{D} \right)^N =  \left(1-\frac{1}{N}\right)^{N-1},$}
\label{equ:opt-R-framed}
\eee
which also applies to the maximum system throughput for the case of $D=1$. Thus,
\eqref{equ:opt-R-framed} is the maximum system throughput for any $N$ and any $D$ such that $N > D$.
Note that when $N>D$, the maximum system throughput of framed slotted ALOHA is the same for any $D \ge 1$.

When $N \le D$, $p^*=1$ and the maximum system throughput is
\bee
& R^*_{\textsf{framed}} (D,N)=R_{\textsf{framed}} (D,N, p^*)  = R_{\textsf{framed}} (D,N, 1) = \frac{N}{D-1} \cdot \left(\frac{D-1}{D}\right)^N.
\label{equ:opt-R-framed-D-larger-than-N}
\eee
Note that by convention
we assume that \eqref{equ:opt-R-framed-D-larger-than-N} includes
the case of $N=D=1$, i.e., $\frac{N}{D-1} \cdot \left(\frac{D-1}{D}\right)^N = \frac{N}{D-1} \cdot \frac{D-1}{D} = \frac{N}{D}=1.$
Thus, \eqref{equ:opt-R-framed-D-larger-than-N} works for any $D \ge 1$.

Clearly, we also have
\be
\lim_{N \to \infty} R^*_{\textsf{framed}} (D,N) = \lim_{N \to \infty}  \left(1-\frac{1}{N}\right)^{N-1} = \frac{1}{e}.
\label{equ:opt-R-converge-to-1-over-e-framed}
\ee

Thus, framed slotted ALOHA has the same asymptotic system throughput as that of $p$-constant slotted ALOHA
and $p$-dynamic slotted ALOHA.
However, there are several advantages of the results of framed slotted ALOHA.
First, for any finite $D$ and $N$, we have the closed-form optimal retransmission probability in \eqref{equ:opt-p-framed},
and the closed-form maximum system throughput as shown in \eqref{equ:opt-R-framed} and \eqref{equ:opt-R-framed-D-larger-than-N}.
However, in our $p$-constant slotted ALOHA, we only have an algorithm to compute $R(D,N,p)$ based on which we
can numerically get the optimal retransmission probability and maximum system throughput,
and in our $p$-dynamic slotted ALOHA, though we have a closed-form optimal retransmission probability,
we still need to run an algorithm to compute the maximum system throughput.
Second, in both $p$-constant and $p$-dynamic slotted ALOHA, the acknowledge information is required to inform the
stations whether their transmissions are successful or not. However,
in the framed slotted ALOHA, since each station will only transmit at most once in a frame, and all packets expire before the end of their frames,
such acknowledge information is not required, which simplifies the system design.


\section{A Reinforcement-Learning-Based Approach to Go Beyond $1/e$} \label{sec:rl-algorithm}

Previous sections proved that three different delay-constrained slotted ALOHA schemes can only achieve the asymptotic timely throughput of $1/e$.
The fundamental reason is that for finite $D$, the number of active stations at any slot goes to infinity as $N \to \infty$ and all active stations will join the competition with the same probability
under ALOHA-based schemes. Thus, the maximum average number of packets delivered successfully in any slot is $1/e$.
Therefore, the asymptotic timely throughput cannot go beyond $1/e$. In order to go beyond $1/e$, we have to reduce the competition level.
The competition level can be effectively reduced by enabling active stations to cooperate with each other but still in a distributed manner without coordination.
So we design a new less-aggressive random access scheme through reinforcement learning to achieve mutual cooperation among active stations.
In this section, we will thoroughly introduce our proposed reinforcement-learning-based random access scheme, called RLRA-DC, which achieves much higher system timely throughput than $1/e$.

Reinforcement learning has been popularly deployed in lots of decision problems in communication systems.
In the paradigm of reinforcement learning, an agent can learn a successful strategy which aims at optimizing objective function by trial-and-error iteration with the environment constantly \cite{sutton2018reinforcement}.
Under our proposed random access scheme RLRA-DC, each station cannot communication with other stations so that it does not know any information about others.
However, they can interact with the AP to infer the information of other stations.
Specifically, at the end of a slot, the AP will broadcast an acknowledgement (ACK) to each station if it decodes a packet successfully,
broadcast a negative-acknowledgement (NACK) if it receives at least two packets but does not decode it successfully (due to channel collision),
and broadcast nothing if it does not receive any packet in this slot.
The idea is borrowed from \cite{yiding2019deep}.

Normally, reinforcement learning approach is characterized by state, action and reward function.
In our RLRA-DC, the state of Station $i$ at slot $t$ is defined as
\be
s_{t,i} \triangleq (l_{t,i}, o_{t,i}), \label{equ:State}
\ee
where $l_{t,i}$ is the lead time  \cite{deng2017timely} of the non-delivered packet (if any) of Station $i$ at slot $t$,
\be
l_{t,i} =
\left\{
  \begin{array}{ll}
    k, & \hbox{if Station $i$ has a packet at slot $t$, which will expire in $k$ slots;} \\
    0, & \hbox{otherwise.}
  \end{array}
\right.
\nnb
\ee
and $o_{t,i} \in \{\textsf{BUSY}, \textsf{SUCCESSFUL}, \textsf{IDLE}, \textsf{FAILED}\}$ is the channel observation of Station $i$ at slot $t$.
Specifically, channel observation $o_{t,i}=\textsf{BUSY}$ means that
Station $i$ does not transmit a packet but receives an ACK from the AP in slot $t-1$,
indicating that
there is only one Station $j$ ($j \neq i$) transmitting a packet in slot $t-1$.
Channel observation $o_{t,i}=\textsf{SUCCESSFUL}$ means that Station $i$ transmits a packet and receives an ACK from the AP in slot $t-1$, indicating that
only Station $i$ transmits a packet in slot $t-1$.
Channel observation $o_{t,i}=\textsf{IDLE}$ means that Station $i$ receive nothing from the AP at the end of slot $t-1$, indicating that there is no station transmitting a packet in slot $t-1$.
Channel observation $o_{t,i}=\textsf{FAILED}$ means that Station $i$ receives an NACK from the AP at the end of slot $t-1$,
indicating that there is at least two stations transmitting packets in slot $t-1$ and a channel collision occurs.
We remark that our model for channel observation is the same as \cite{yiding2019deep}.
By convention, we set $o_{1,i}=\textsf{IDLE}$.

We define $\mathcal{S}_{t,i}$ as 
the set of all possible system states of Station $i$ at slot $t$.
We further define $\mathcal{S}_i$ as the set of all possible system states of Station $i$, i.e., $\mathcal{S}_i \triangleq \cup_{t=1}^{\infty} \mathcal{S}_{t,i}$.
We remark that $\mathcal{S}_{t,i}$ may not be equal to $\mathcal{S}_{t',i}$ when $t \neq t'$. For example,
for any $D$, $o_{1,i}=\textsf{IDLE}$ but $o_{2,i}$ could be any one in $\{\textsf{BUSY}, \textsf{SUCCESSFUL}, \textsf{IDLE}, \textsf{FAILED}\}$.
In addition, for any $D \ge 2$, $l_{2,i}$ can only be 0 or $D-1$. In other words, it is possible that $s_{t,i}$ cannot take any value in
$\mathcal{S}_i$ for some slot $t$. For $D=1$, $l_{t,i}=1$ for any $t$ and thus we have $\left|\mathcal{S}_i \right|=4$. For any $D \ge 2$, we have $\left|\mathcal{S}_i \right|=4(D+1)$.

At slot $t$, the action of Station $i$ is denoted by $a_{t,i}$. Similar to \cite{yiding2019deep}, the action space of Station $i$ is defined as $\mathcal{A}_i \triangleq \{\textsf{TRANSMIT}, \textsf{WAIT}\}$.
Action $a_{t,i}=\textsf{TRANSMIT}$ means that Station $i$ transmits its packet at slot $t$, while $a_{t,i}=\textsf{WAIT}$ means that
Station $i$ does not transmit a packet at slot $t$.

We define the reward function $r_i(s_{t,i},a_{t,i})$ of Station $i$ as
\bee
r_i(s_{t,i},a_{t,i}) \triangleq 1_{\left\{o_{t,i} \in \{ \textsf{BUSY}, \textsf{SUCCESSFUL}\} \right \}}, \forall s_{t,i} \in \mathcal{S}_i, a_{t,i} \in \mathcal{A}_i,
\label{equ:reward function}
\eee
where $1_{\{\cdot\}}$ is the indicator function. Note that $o_{t,i}=\textsf{BUSY}$
means that Station $j$ ($j\neq i$) transmits a packet successfully in slot $t-1$,
and $o_{t,i}= \textsf{SUCCESSFUL}$ means that Station $i$ transmits a packet successfully in slot $t-1$.
Thus $r_i(s_{t,i},a_{t,i})=1$ if the system transmits a packet successfully in slot $t-1$.
Note that we model the reward with ``delay of gratification".

\begin{algorithm}[t]
 \caption{{R}einforcement-{L}earning-based {R}andom {A}ccess scheme for {D}elay-{C}onstrained communications (RLRA-DC) of station $i$}
 \label{alg:rl}
\begin{algorithmic}[1]
    \STATE Initialize relative value $Q_i(s, a) = 0, \forall s \in {\mathcal{S}_i}, \forall a \in \mathcal{A}_i$
    \STATE Initialize average reward $\rho_i = 0$
    \STATE Set learning rates $\alpha_i=0.01, \beta_i=0.01$
    \STATE Observe the initial system state $s_{1,i}$
    \FOR{$t = 1, 2, \cdots$}
	    \IF{$t \le 4D$}
	    	\STATE Station $i$ adopt $p$-constant ALOHA protocol with transmission probability $p_{t,i} = 1/(2N)$ where $N$ is the number of stations. Namely,
    	    \[
		    a_{t,i}=
		    \left\{
		      \begin{array}{ll}
		        \textsf{TRANSMIT}, & \hbox{with prob. $p_{t,i}$;} \\
		        \textsf{WAIT}, & \hbox{with prob. $1-p_{t,i}$,}
		      \end{array}
		    \right.
		    \]
    	\ELSE
    		\STATE $a_{t,i} = \arg \max_{a \in \mathcal{A}_i} Q_i(s_{t,i}, a)$
	    \ENDIF

	    \STATE Observe reward $r_i(s_{t,i}, a_{t,i})$ and state $s_{t+1, i}$
	    \STATE Update relative value as follow,
	    		\bee
				Q_i(s_{t,i}, a_{t,i}) & \leftarrow Q_i(s_{t,i}, a_{t,i}) + \alpha_i \big( r_i(s_{t,i}, a_{t,i})  + \max_{a}Q_i(s_{t+1, i}, a) - Q_i(s_{t,i}, a_{t,i}) - \rho_i \big) \nnb
				\eee
	    \STATE Update average reward as follow,
	    		\bee
	    		\rho_i & \leftarrow \rho_i + \beta_i\big( r_i(s_{t,i}, a_{t,i}) + \max_{a}Q_i(s_{t+1, i}, a) - Q_i(s_{t,i}, a_{t,i}) - \rho_i \big) \nnb
	    		\eee
    \ENDFOR

\end{algorithmic}
\end{algorithm}

The most-widely used reinforcement learning algorithm is Q-learning which applies to the discounted-reward case \cite{sutton2018reinforcement}.
However, the objective in our problem, i.e., the system timely throughput, is an average reward as defined in \eqref{equ:def-R}.
Thus, motivated by \cite{danzhou2021reinforcement}, we think that Q-learning is less suitable to R-learning, a less-popular reinforcement learning algorithm,
which applies to the average-reward case \cite{schwartz1993reinforcement, singh1994reinforcement,sutton2018reinforcement}.


In this paper, we adopt the variants of R-learning in \cite[Algorithm 3]{singh1994reinforcement} and \cite[Figure 11.2]{sutton2018reinforcement}.
For any Station $i$, the iterations of relative value $Q_i(s_{t,i}, a_{t,i})$ and average reward $\rho$ are as follows,
\bee
Q_i(s_{t,i}, a_{t,i}) & \leftarrow Q_i(s_{t,i}, a_{t,i}) + \alpha_i \big( r_i(s_{t,i}, a_{t,i})  + \max_{a \in \mathcal{A}_i}Q_i(s_{t+1, i}, a) - Q_i(s_{t,i}, a_{t,i}) - \rho_i \big), \label{equ:upgrade_Q}\\
\rho_i & \leftarrow \rho_i + \beta_i\big( r_i(s_{t,i}, a_{t,i}) + \max_{a \in \mathcal{A}_i }Q_i(s_{t+1, i}, a) - Q_i(s_{t,i}, a_{t,i}) - \rho_i \big), \label{equ:upgrade_rho}
\eee
where $\alpha_i \in (0,1]$ and $\beta_i \in (0,1]$ are learning rates, $\rho_i$ approximates the state-independent
average reward for the iteratively updated policy $\pi$, i.e.,
\be
\rho_i \approx \lim_{T \to \infty} \mathbb{E}^{\pi} \left[ \frac{\sum_{t=1}^T r_i(s_{t,i},a_{t,i})}{T} \right],
\label{equ:rho-approx-R-learning}
\ee
and Q-function $Q_i(s_{t,i}, a_{t,i})$ approximates the state-dependent cumulative reward difference (called relative value in \cite{singh1994reinforcement,sutton2018reinforcement}) for the iteratively updated policy $\pi$, i.e.,
\be
Q_i(s_{t,i}=s, a_{t,i}=a)  \approx  \mathbb{E}^{\pi}  \left[ \left. \sum_{\tau=t}^{\infty}  \left[r_i(s_{\tau,i},a_{\tau,i}) - \rho_i \right] \right| s_{t,i} = s, a_{t,i}=a \right].
\label{equ:Q-approx-R-learning}
\ee
Specifically, our proposed algorithm is called {R}einforcement-{L}earning-based {R}andom {A}ccess scheme for {D}elay-{C}onstrained communications (RLRA-DC) of station $i$ for a given number of station, i.e., $N$, as shown in Algorithm \ref{alg:rl}.

As we can see in Algorithm \ref{alg:rl}, in the first $4D$ slots (which is a short period), each station adopts $p$-constant slotted ALOHA protocol with transmission probability $1/(2N)$ (which is a small value) to initialize its Q-table. The goal of this procedure is to create some heterogeneity of initialized policies of all stations.
Furthermore, the small transmission probability $1/(2N)$ reduces the collision probability during the initialization period (from slot 1 to slot $4D$).
In other words, we essentially pre-allocate a part of spectrum resources to a part of stations in the first $4D$ slots, which benefits the convergence of the algorithm.
And the rest of spectrum resources will be allocated judiciously in later slots through the interaction between stations.
We will show the superior performance of Algorithm \ref{alg:rl} in Fig. \ref{fig:RL-result}.
\begin{algorithm}[t]
 \caption{The method of Station $i$ to estimate the number of stations, i.e., $N$}
 \label{alg:prediction}
\begin{algorithmic}[1]
	\STATE Initialize a constant $p = \frac{1}{10}$
	\FOR {$k = 1,2, \cdots, 100$}

		\FOR {$t = 1,2, \cdots, 100$}
			\STATE Station $i$ adopts $p$-constant ALOHA protocol with transmission probability $p_t= \frac{p}{k}$
		\ENDFOR

		\STATE Calculate the empirical timely throughput $R(k)$ from slot $100(k-1) + 1$ to $100k$
	\ENDFOR

	\STATE The estimation of station $i$ for the number of stations is $N = \frac{\arg \max_{k\in \{1,2,\cdots, 100\}} R(k)}{p}$

\end{algorithmic}
\end{algorithm}


Similar to $p$-constant slotted ALOHA and framed slotted ALOHA , Algorithm \ref{alg:rl} again requires that $N$ is known a priori, which is sometimes difficult to be obtained in practice.
Hence, we also propose a simple but practical method to estimate $N$.
The idea is inspired by Theorem \ref{thm:opt-p-1-N} in ALOHA system, which shows that the optimal transmission probability behaves asymptotically as $1/N$.
We assume that the number of stations $N$ in the network is not larger than 1,000.
The algorithm to estimate the number of stations $N$ is shown in Algorithm \ref{alg:prediction},
whose effectiveness will be demonstrated in Section \ref{sec:simulation}.

\section{Simulation} \label{sec:simulation}
In this section, we first confirm our theoretical analysis by simulations for the three delay-constrained slotted ALOHA schemes in Section \ref{subsec:sim-ALOHA},
and then demonstrate the asymptotic performance of ALOHA-based schemes and RLRA-DC in Section \ref{subsec:sim-asy-ALOHA} and Section \ref{subsec:sim-asy-RLRA}, respectively.

\subsection{Confirmation of Theoretical Analysis for System Throughput} \label{subsec:sim-ALOHA}
\begin{figure*}[t]
  \centering
  \includegraphics[width=\linewidth]{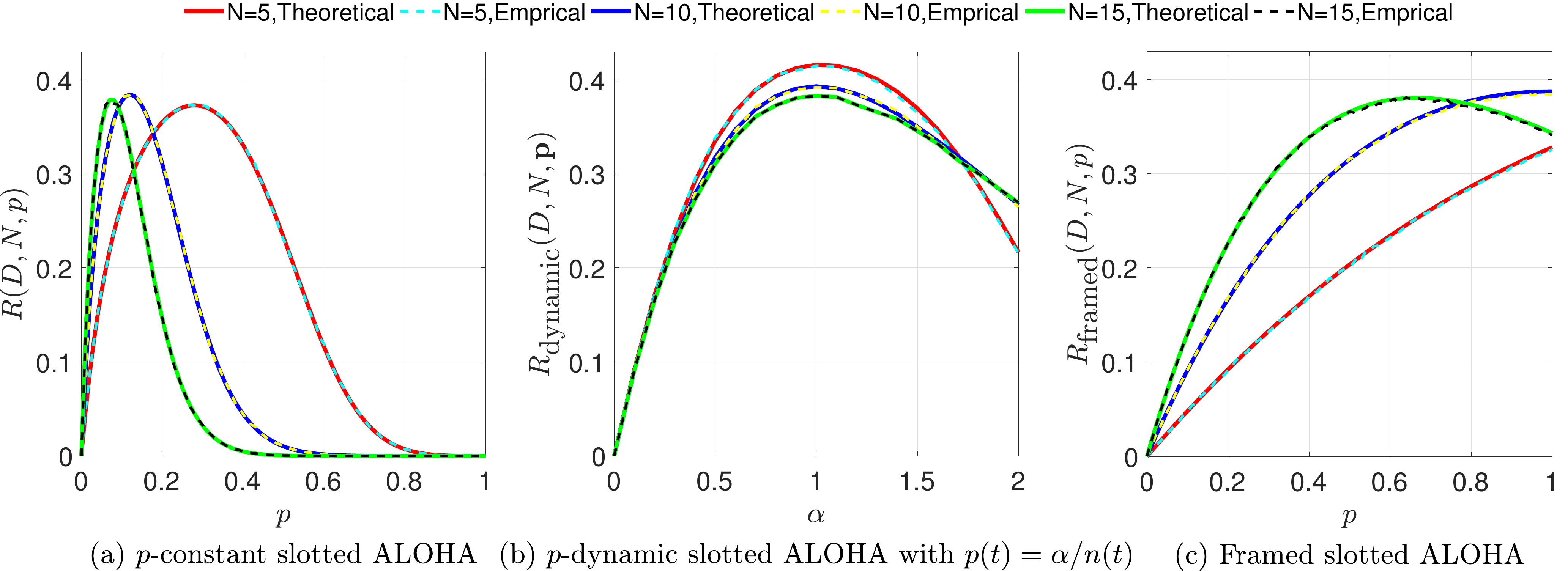}\\
  \caption{The system throughput of $p$-constant slotted ALOHA, $p$-dynamic slotted ALOHA, and
framed slotted ALOHA with $D=10$. }\label{fig:verification} \vspace{-0.8cm}
\end{figure*}

We first confirm our theoretical analysis for $p$-constant slotted ALOHA, $p$-dynamic slotted ALOHA,
and framed slotted ALOHA. For $p$-constant slotted ALOHA (resp. $p$-dynamic slotted ALOHA),
we can use our algorithm in Section~\ref{sec:algorithm} to compute the system throughput $R(D,N,p)$
(resp. $R_{\textsf{dynamic}}(D,N,\bm{p})$) for any given hard delay $D$, number of stations $N$,
retransmission probability $p$ (resp. retransmission policy $\bm{p}=(p(n,t): n=1,2,\cdots,N, t=1,2,\cdots,D)$).
In this simulation, we consider the dynamic retransmission policy
$p(t)=\alpha/n(t)$ where $\alpha$ is a constant ranging from 0 to 2, and $n(t)$ is the number
of active stations at the beginning of slot $t$.
The reason that we set $p(t)=\alpha/n(t)$ is because
we aim to verify that the optimal system throughput is achieved when $\alpha=1$ and also to check the variation
of the system throughput due to $\alpha$.
For framed slotted ALOHA, we have a closed form
for the system throughput $R_{\textsf{framed}}(D,N,p)$ as shown in
\eqref{equ:opt-R-framed} and \eqref{equ:opt-R-framed-D-larger-than-N}.
Thus, for all three schemes, we can get the theoretical system throughput.
To verify the correctness, we further simulate a real system for the three schemes
and obtain the empirical system throughput.
We fix hard delay $D=10$ and consider $N=5,10,$ and $15$
and simulate these three schemes for 10,000 frames (of in total 100,000 slots) to obtain the empirical system throughput.
The results are shown in Fig.~\ref{fig:verification}.

For all three ALOHA-based schemes, as expected, we can see that the theoretical system throughput is the same as the empirical system throughput,
which confirms  our theoretical derivation for system throughput for all three schemes.
From the middle figure in Fig.~\ref{fig:verification} for $p$-dynamic slotted ALOHA, we can see
that the system throughput is indeed maximized at $\alpha=1$, which confirms the optimal
retransmission policy in \eqref{equ:opt-p-dynamic-3}. From the right-hand-side figure
in Fig.~\ref{fig:verification} for framed slotted ALOHA, we can see that the optimal retransmission probability
is $\min\{D/N,1\}$, which confirms \eqref{equ:opt-p-framed}.

\subsection{Asymptotic Performance of ALOHA-based Schemes} \label{subsec:sim-asy-ALOHA}
\begin{figure}
\subfigure[Maximum system throughput of three ALOHA-based schemes.]{\includegraphics[width=0.3\linewidth]{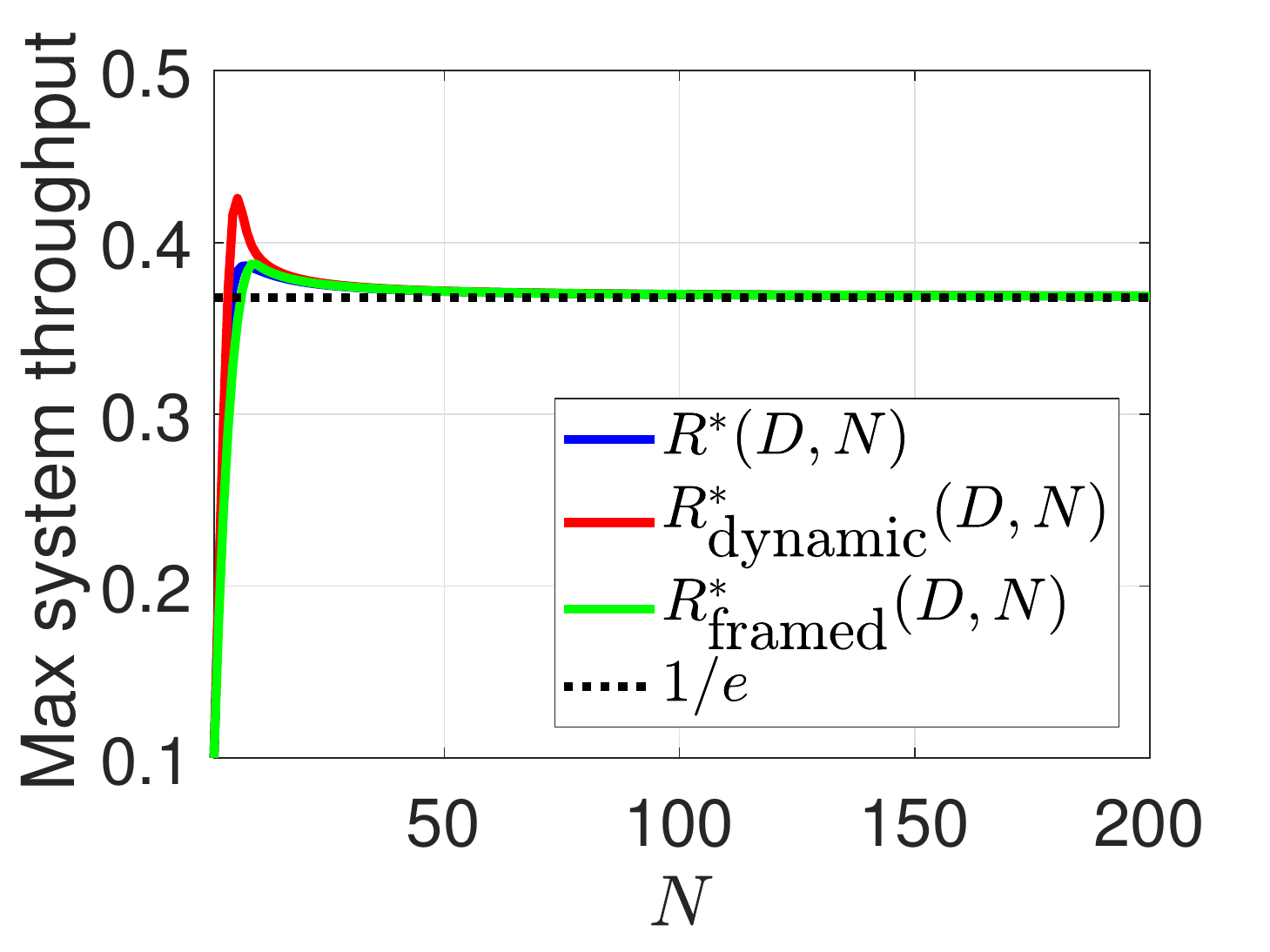}\label{fig:asymptotic-opt-R}}
\subfigure[A Small range of $N$ (from 1 to 15) for Fig.~\ref{fig:asymptotic-opt-R}.]{\includegraphics[width=0.3\linewidth]{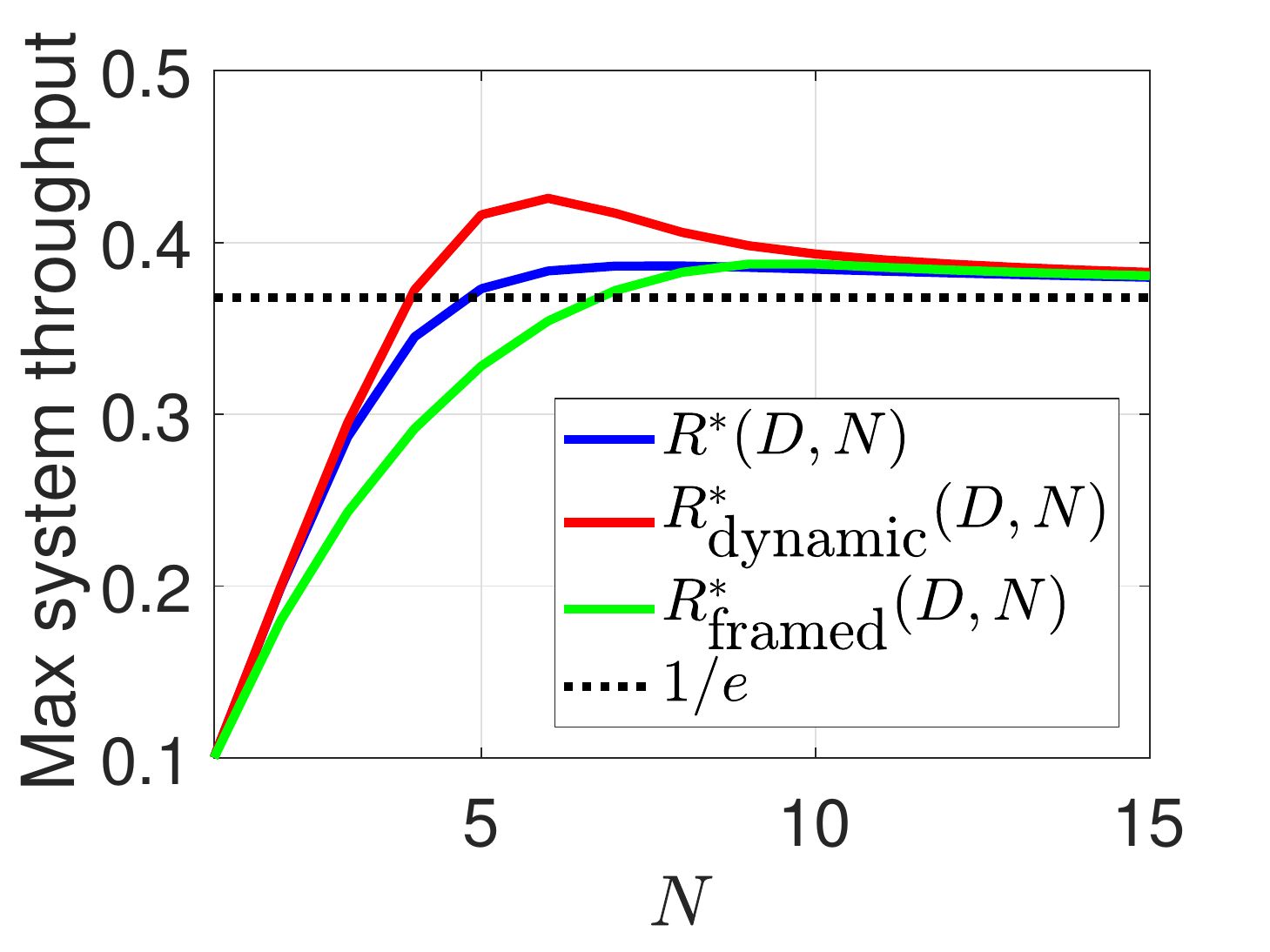}\label{fig:asymptotic-opt-R-small-range}}
\subfigure[Optimal retransmission prob. of $p$-constant slotted ALOHA.]{\includegraphics[width=0.3\linewidth]{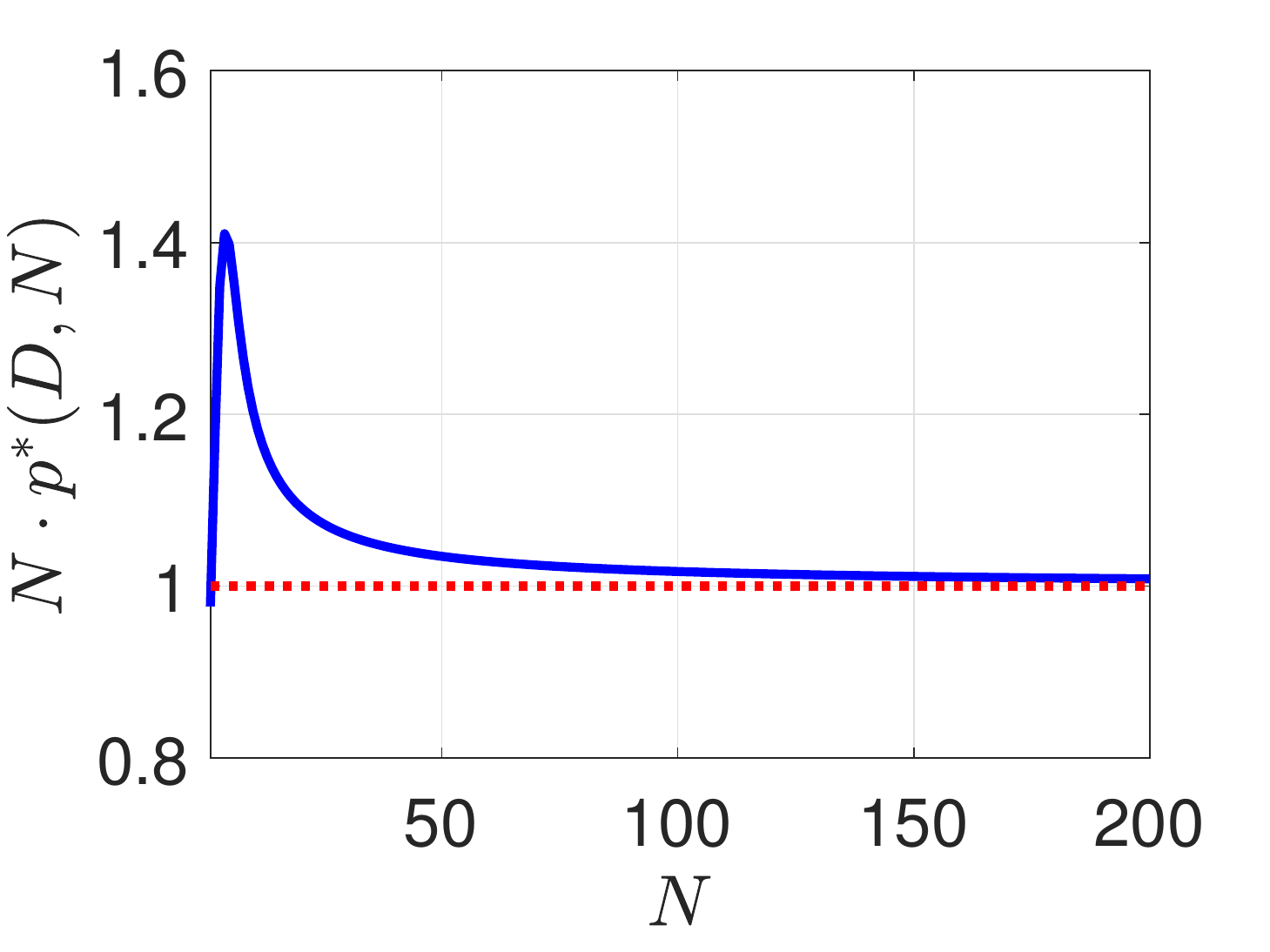}\label{fig:asymptotic-opt-p}}
\caption{The asymptotic performance of the maximum system throughput for three ALOHA-based
schemes and the asymptotic performance of the optimal retransmission probability of $p$-constant slotted ALOHA
when $D=10$. \label{fig:asymptotic-opt-R-opt-p}}
\end{figure}

We consider $D=10$ and show the asymptotic performance of maximum system throughput of three schemes
and the asymptotic performance of the  optimal retransmission probability of $p$-constant slotted ALOHA
in Fig.~\ref{fig:asymptotic-opt-R-opt-p}.
First, from Fig.~\ref{fig:asymptotic-opt-R}, we can see that indeed the maximum system throughput of all three schemes converges to $1/e$,
which confirms Theorem~\ref{thm:converge-to-1-over-e}, Theorem~\ref{thm:opt-R-converge-to-1-over-e-dynamic},
and \eqref{equ:opt-R-converge-to-1-over-e-framed}. Second, to better compare three schemes,
Fig.~\ref{fig:asymptotic-opt-R-small-range} shows the maximum system throughput of all three schemes when $N$ ranges from 1 to 15.
We can see that in terms of maximum system throughput,
$p$-dynamic slotted ALOHA has the best performance for any $N$;
$p$-constant slotted ALOHA is better than framed slotted ALOHA when $N \le 8$ but worse than framed slotted ALOHA when $N \ge 9$ (though with very close performance).
Finally, from Fig.~\ref{fig:asymptotic-opt-p}, we can see that indeed
the optimal retransmission probability of $p$-constant slotted ALOHA behaves asymptotically as $1/N$, which verifies Theorem~\ref{thm:opt-p-1-N}.


\subsection{Asymptotic Performance of RLRA-DC} \label{subsec:sim-asy-RLRA}


\begin{figure}
\subfigure[The timely throughput of the 10-user case.]{\includegraphics[width=0.48\linewidth]{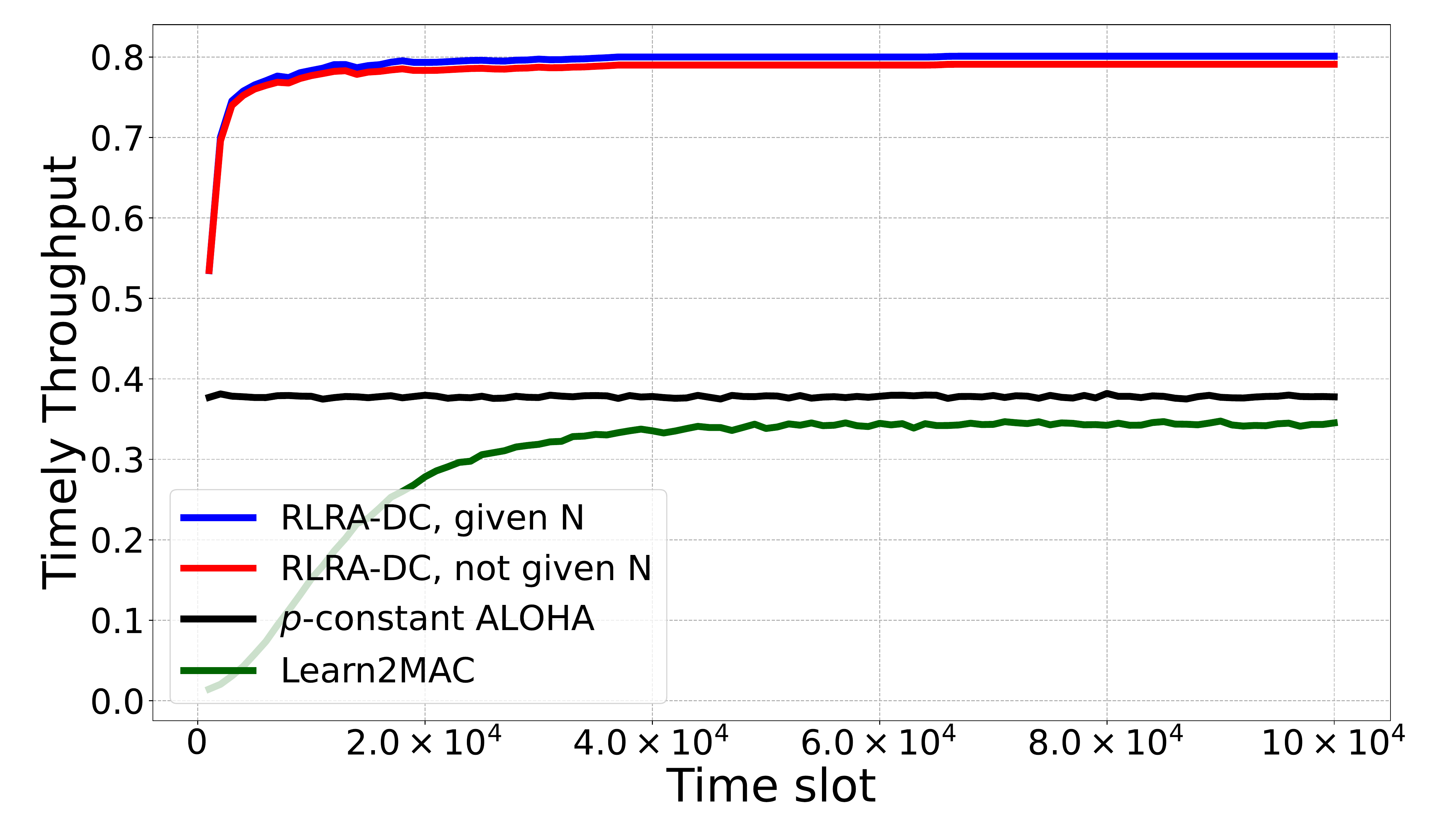}\label{fig:N=10_D=10}}
\subfigure[The timely throughput of the 50-user case.]{\includegraphics[width=0.48\linewidth]{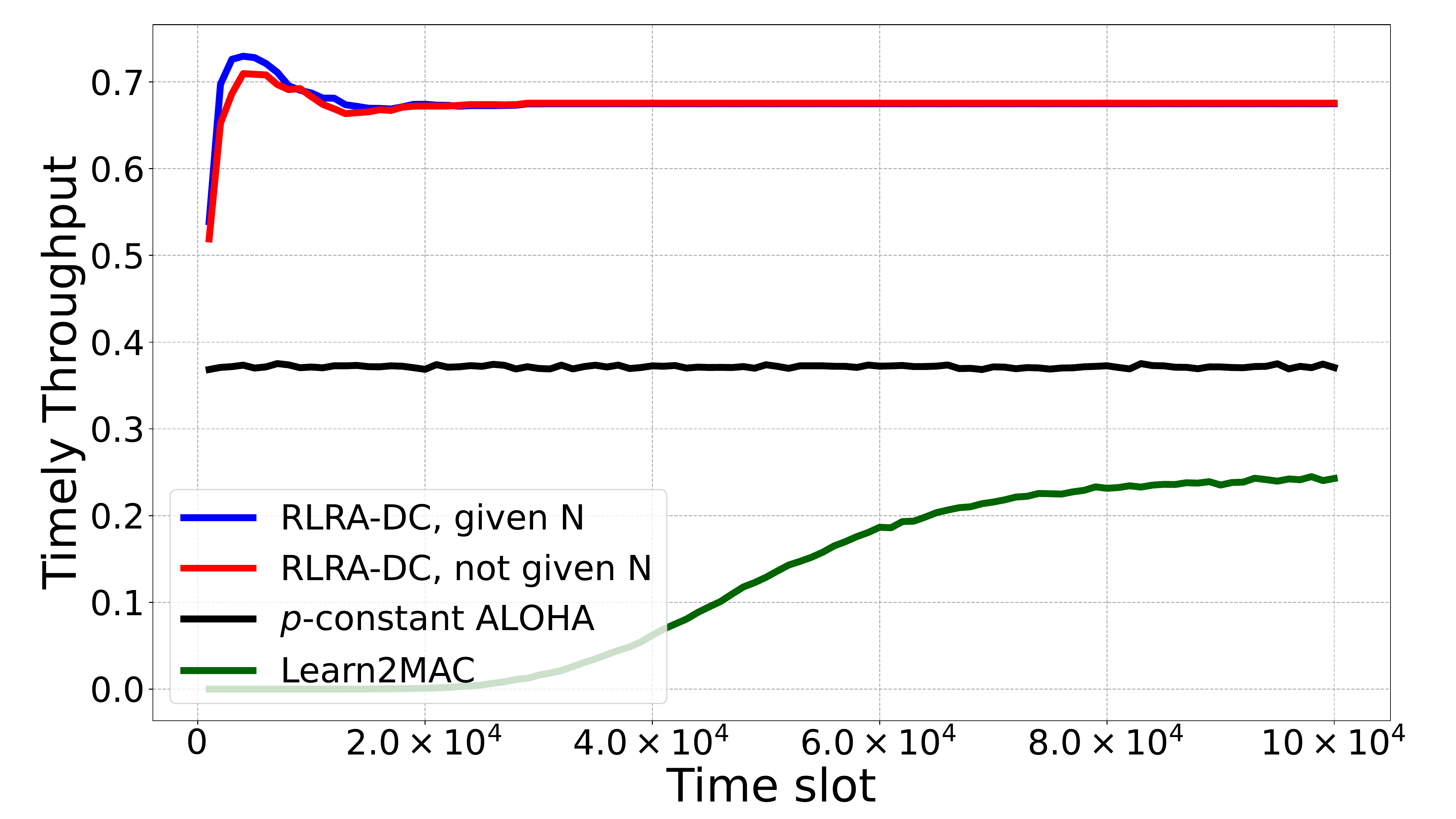}\label{fig:N=50_D=10}}
\\
\subfigure[The timely throughput of the 100-user case.]{\includegraphics[width=0.48\linewidth]{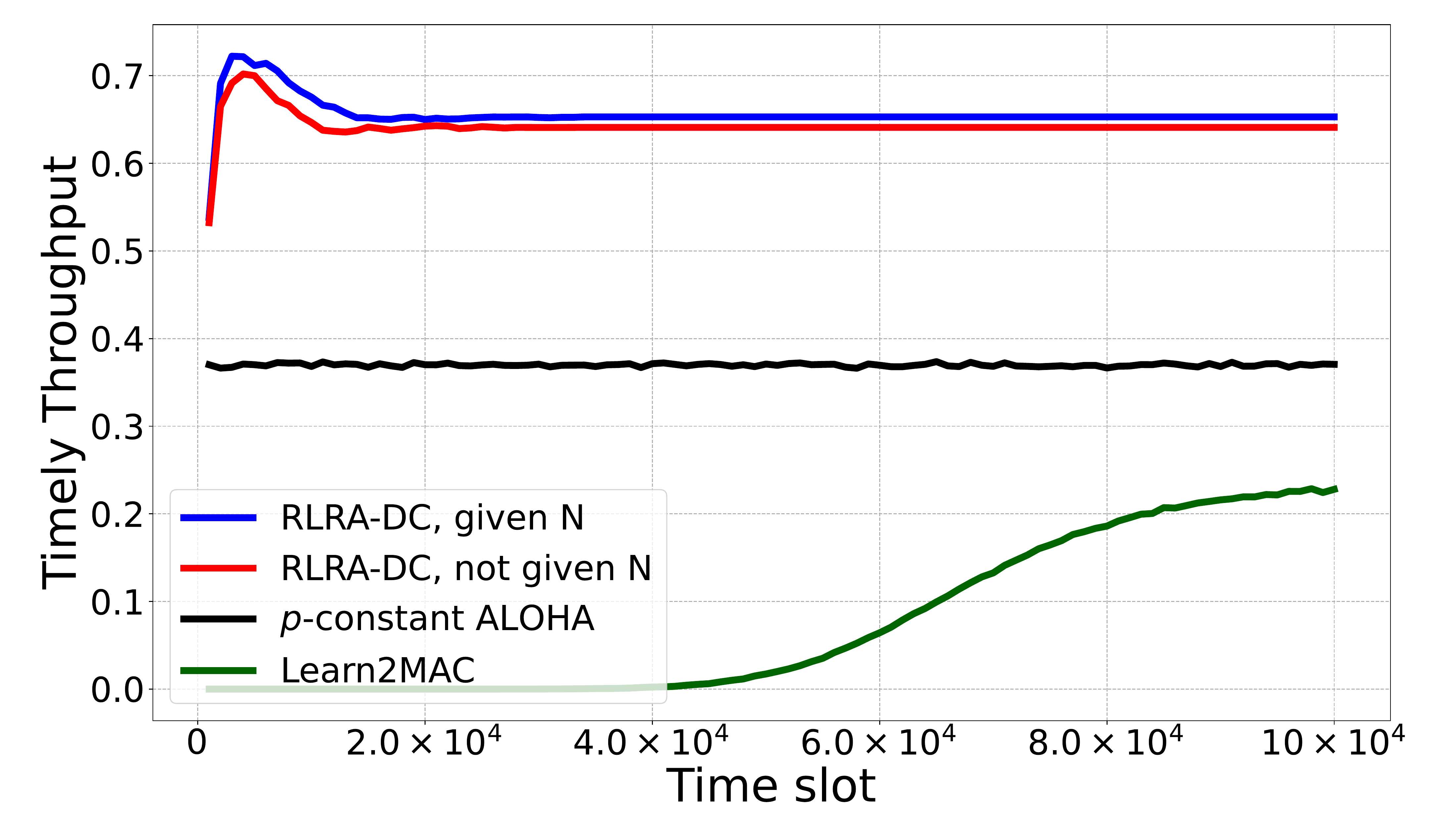}\label{fig:N=100_D=10}}
\subfigure[The timely throughput of the 1000-user case.]{\includegraphics[width=0.48\linewidth]{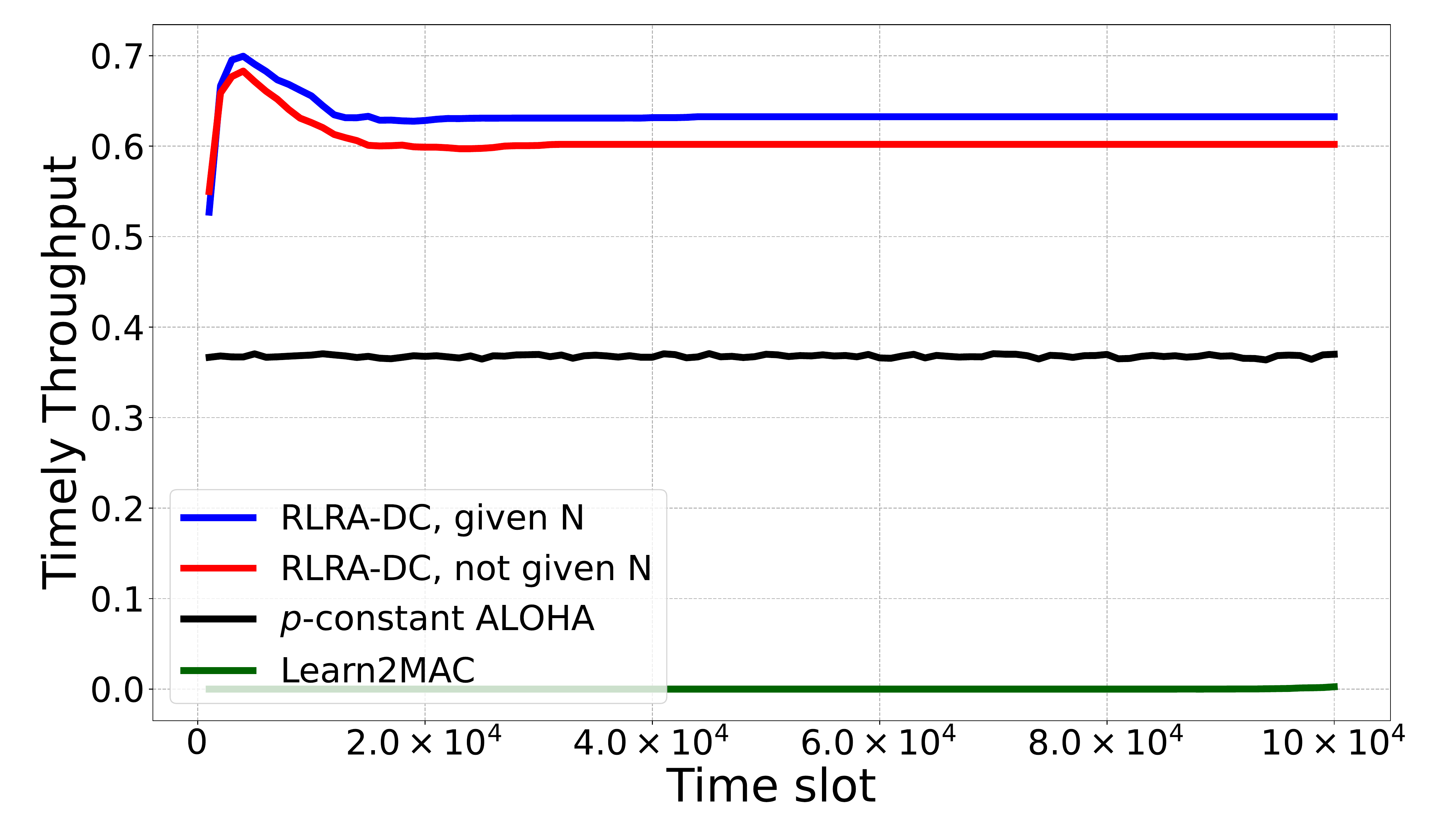}\label{fig:N=1000_D=10}}
\caption{The system timely throughput of four random access schemes when $D=10$ and $N=$ 10, 50, 100, and 1,000, respectively. \label{fig:RL-result}}
\end{figure}

Finally, we demonstrate the superior performance of our proposed RLRA-DC described in Section \ref{sec:rl-algorithm}.
We set the deadline $D=10$, and the number of stations $N$  to be 10, 50, 100, and 1,000, respectively.
For each $N$, we independently run 100 instances with different random seeds ranging from 1 to 100.
In each instance, we run 100,000 slots and get the system timely throughput.
We then average the system timely throughput of these 100 instances as the result.
The baselines are the online-learning-based algorithm called Learn2MAC \cite{destounis2019learn2mac}\footnote{Note that we have found a typo in \cite{destounis2019learn2mac} and corrected it.
Specifically, The equation of step 5 in Algorithm 1 in \cite{destounis2019learn2mac} should be
$p^i_k(t) = \frac{p^j_k(t-1)e^{\alpha v_k^i}}{\sum^d_{j=1}p^j_k(t-1)e^{\alpha v_k^i}}$
instead of
$p^i_k(t) = \frac{p^j_k(t-1)e^{-\alpha v_k^i}}{\sum^d_{j=1}p^j_k(t-1)e^{-\alpha v_k^i}}$.} and $p$-constant slotted ALOHA.
The system timely throughput of these random access schemes are shown in Fig. \ref{fig:RL-result}.

From Fig. \ref{fig:RL-result}, we have the following four observations.
First, our RLRA-DC with given $N$ always achieves the highest system timely throughput in each case, which is much larger than $1/e$.
Second, when $N$ is not given, by using our proposed method to estimate the number of station in Algorithm \ref{alg:prediction}, the system timely throughput of RLRA-DC is very close to that of the case when $N$ is given.
It demonstrates that our estimation method in Algorithm \ref{alg:prediction} is efficient despite its simplicity.
Third, the system timely throughput of $p$-constant slotted ALOHA converges to $1/e$ which is consistent with our previous sections.
And the system timely throughput of Learn2MAC is even lower than that of $p$-constant slotted ALOHA.
The reason is that Learn2MAC only works in the less-congested scenarios, as mentioned in \cite{destounis2019learn2mac}.
Finally,
as $N$ increases, the collision probability also increases, which will reduce the system timely throughput.
Nevertheless, even when the number of stations $N$ increases to $1,000$, the system timely throughput of our RLRA-DC is $0.6$, which is much higher than $1/e$.
{Our independent simulation shows that the system timely throughput of up to 10,000 stations is also very close to 0.6, much higher than $1/e$}.


Thus, Fig. \ref{fig:RL-result} shows the superior performance of our proposed R-learning-based random access scheme RLRA-DC.
Its system timely throughput can be much larger than $1/e$, which is the ceiling of delay-constrained ALOHA schemes.

\section{Conclusion} \label{sec:conclusion}
In this paper, we have firstly analyzed the asymptotic performance of three
delay-constrained slotted ALOHA schemes under the
frame-synchronized traffic pattern. We have proved that
the maximum system throughput of all three schemes
converges to $1/e$ when the number of stations goes to infinity.
We have also characterized the optimal retransmission probability of
three schemes. Then we have proposed a reinforcement-learning-based random access scheme for delay-constrained communications called RLRA-DC,
whose the system timely throughput is much larger than $1/e$.
In the future, to capture more practical scenarios,
it is interesting to further study the delay-constrained communication
under non-frame-synchronized traffic patterns.
In addition, it is also important to investigate
the performance of other medium access schemes, e.g., CSMA,
under delay-constrained setting.



\bibliographystyle{IEEEtran}
\bibliography{ref}

\newpage

\appendix

\subsection{Proof of Lemma~\ref{lem:a-key-lemma}} \label{app:proof-of-lem-a-key-lemma}
We consider two cases: $\delta < \infty$, and $\delta = \infty$.

\emph{Case I: $\delta < \infty$.}
In this case, due to \eqref{equ:p(N)-to-0} and \eqref{equ:Np(N)-to-1}, we have
\bee
& \lim\limits_{k \to \infty}  (N_k-A) p(N_k) \left[1- p(N_k)\right]^{N_k-B}  = \lim\limits_{k \to \infty} (N_k-A) p(N_k) \cdot \lim\limits_{k \to \infty} \left[1- p(N_k)\right]^{N_k}  \cdot \lim\limits_{k \to \infty} \frac{1}{[1-p(N_k)]^{B}} \nnb \\
& =  \delta \cdot  \lim_{k \to \infty} \exp \left\{ N_k \ln \left[1-p(N_k) \right] \right\} \cdot 1   = \delta \cdot \exp \left\{\lim_{k \to \infty}  N_k \ln \left[1-p(N_k) \right] \right\} \nnb \\
& = \delta \cdot \exp \left\{\lim_{k \to \infty}  N_k p(N_k) \cdot \frac{\ln \left[1-p(N_k) \right]}{p(N_k)} \right\}  = \delta \cdot \exp \left\{ \lim_{k \to \infty}  N_k p(N_k) \cdot \lim_{k \to \infty}  \frac{\ln \left[1-p(N_k) \right]}{p(N_k)} \right\} \nnb \\
& \overset{(\text{let } u=p(N_k))}{=} \delta \cdot \exp \left\{ \delta \cdot \lim_{u \to 0}  \frac{\ln (1-u)}{u} \right\}
= \delta \cdot \exp \left\{ \delta \cdot \lim_{u \to 0}  \frac{-\frac{1}{1-u}}{1} \right\} \quad (\text{by L'Hospital's Rule}) \nnb \\
& = \delta \cdot \exp(-\delta) = \frac{\delta}{e^{\delta}}. \nnb
\eee
Thus, \eqref{equ:limit-is-delta-over-exp-delta} holds when $\delta < \infty$.

\emph{Case II: $\delta = \infty$}. In this case, we have
\bee
\resizebox{0.92\linewidth}{!}{$0  \le  N_k p(N_k) [1-p(N_k)]^{N_k}  =  N_k p(N_k) \left\{ \left[1-p(N_k)\right]^{\frac{1}{p(N_k)}} \right\}^{N_kp(N_k)}
\le N_k p(N_k) \left( \frac{1}{e} \right)^{N_kp(N_k)} = \frac{N_k p(N_k)}{e^{N_k p(N_k)}}$},
\label{equ:squeeze-np(N)-inequ}
\eee
where the last inequality follows from the fact that
\be
(1-x)^{\frac{1}{x}} < \frac{1}{e}, \forall x \in (0,1]. \nnb
\ee
Since $\lim_{k \to \infty} N_k p(N_k) = \delta = \infty$, by letting $u=N_k p(N_k)$, we have
\be
\lim_{k \to \infty} \frac{N_k p(N_k)}{e^{N_k p(N_k)}} = \lim_{u \to \infty} \frac{u}{e^u} = 0. \nnb
\ee
Thus, by applying the squeeze theorem to \eqref{equ:squeeze-np(N)-inequ}, we have
\be
\lim_{k \to \infty} N_k p(N_k) [1-p(N_k)]^{N_k} = 0.
\label{equ:NpN-to-0}
\ee
Note that \eqref{equ:NpN-to-0} implies
\be
\lim_{k \to \infty}  p(N_k) [1-p(N_k)]^{N_k} = 0.
\label{equ:M1pN-to-0}
\ee
Combining \eqref{equ:NpN-to-0} and \eqref{equ:M1pN-to-0}, we have
\bee
& \lim_{k \to \infty}  (N_k-A) p(N_k) \left[1- p(N_k)\right]^{N_k-B}  = \lim_{k \to \infty}  (N_k-A) p(N_k) \left[1- p(N_k)\right]^{N_k} \cdot \lim_{k \to \infty} \frac{1}{[1-p(N_k)]^{B}} \nnb \\
& = \lim\limits_{k \to \infty}  (N_k-A) p(N_k) \left[1- p(N_k)\right]^{N_k} \cdot 1   = \lim\limits_{k \to \infty} N_k p(N_k) \left[1- p(N_k)\right]^{N_k} - A \cdot \lim\limits_{k \to \infty}  p(N_k) \left[1- p(N_k)\right]^{N_k} \nnb \\
& = 0 - A \cdot 0 = 0 = \frac{\infty}{e^{\infty}} = \frac{\delta}{e^{\delta}}, \nnb
\eee
where the second last equality uses the convention \eqref{equ:convention}.
Thus, \eqref{equ:limit-is-delta-over-exp-delta} holds when $\delta = \infty$.

\emph{Case I} and \emph{Case II} complete the proof.

\subsection{Proof of Lemma~\ref{lem:prob-m-t-limit-exist}} \label{app:proof-of-prob-m-t-limit-exist}
We prove the existence of the limit (i.e., \eqref{equ:P-M-N-t=m-limit-exist}) by induction with respect to $t$.
Clearly, when $t=1$,
\be
P(M_{N,1} = 0) = 1, \quad \forall N. \nnb
\ee
Thus,
\be
\lim_{k \to \infty} P(M_{N_k,1}=0)  = 1 \triangleq \gamma_1^0. \nnb
\ee

For $t \in \{1,2,\cdots, D-1\}$, suppose that $\lim_{k \to \infty} P(M_{N_k,t} = m)$  exists for any $m$, i.e.,
\bee
& \lim_{k \to \infty} P(M_{N_k,t}=m) = \gamma_t^m, \quad  m \in  \{0,1,\cdots, t-1\}. \nnb
\eee

We then consider $t+1$. For $m=0$,
\bee
& P(M_{N_k,t+1} = m) = P(M_{N_k,t+1} = 0)  = \sum_{m'=0}^{t-1} P(M_{N_k,t+1} = 0 | M_{N_k,t} = m')P(M_{N_k,t}=m') \nnb \\
& \resizebox{0.92\linewidth}{!}{$= P(M_{N_k,t+1} = 0 | M_{N_k,t} = 0) P(M_{N_k,t}=0)  = \left\{1 - N_k p(N_k) \left[1-p(N_k)\right]^{N_k-1}\right\}P(M_{N_k,t}=0)$}. \label{equ:P-m-n-k+1=0}
\eee
Since $p(N_k)$ satisfies \eqref{equ:p(N)-to-0} and \eqref{equ:Np(N)-to-1}, according to Lemma~\ref{lem:a-key-lemma} where we set $A=0, B=1$, we have
\be
\lim_{k \to \infty} N_k p(N_k) \left[1-p(N_k)\right]^{N_k-1} = \frac{\delta}{e^{\delta}}. \nnb
\ee
Therefore, taking limit in \eqref{equ:P-m-n-k+1=0}, we have
\bee
\resizebox{0.92\linewidth}{!}{$ \lim\limits_{k \to \infty} P(M_{N_k,t+1}=0)  = \lim\limits_{k \to \infty}  \left\{1 - N_k p(N_k) \left[1-p(N_k)\right]^{N_k-1}\right\} \cdot \lim\limits_{k \to \infty} P(M_{N_k,t}=0)
 = \left(1 - \frac{\delta}{e^\delta} \right) \cdot \gamma_t^0 \triangleq \gamma_{t+1}^0$}. \label{equ:P-m-n-k+1=0-limit}
\eee
Thus, the limit $\lim_{k \to \infty} P(M_{N_k,t+1}=0)$ exists.

Similarly, for any $m \in \{1,2,\cdots, t-1\}$, we have
\bee
& P(M_{N_k,t+1} = m)  = \sum_{m'=0}^{t-1} P(M_{N_k,t+1} = m | M_{N_k,t} = m') P(M_{N_k,t} = m') \nnb \\
& = P(M_{N_k,t+1} = m | M_{N_k,t} = m-1) P(M_{N_k,t}=m-1) + P(M_{N_k,t+1} = m | M_{N_k,t} = m) P(M_{N_k,t}=m) \nnb \\
& \resizebox{0.99\linewidth}{!}{$= [N_k-(m-1)] p(N_k) [1-p(N_k)]^{N_k-(m-1)-1}P(M_{N_k,t}=m-1) + \left\{1-(N_k-m) p(N_k) [1-p(N_k)]^{N_k-m-1}\right\} P(M_{N_k,t}=m).$}\nnb
\eee
According to Lemma~\ref{lem:a-key-lemma}, we have
\be
\lim_{k \to \infty} P(M_{N_k,t+1} = m) = \frac{\delta}{e^\delta} \cdot \gamma_t^{m-1} + \left(1-\frac{\delta}{e^\delta}\right) \gamma_t^{m} \triangleq \gamma_{t+1}^m.\nnb
\ee
Finally, for $m=t$, we have
\bee
P(M_{N_k,t+1} = t)  &= P(M_{N_k,t+1} = t | M_{N_k,t} = t-1) P(M_{N_k,t}=t-1) \nnb \\
& = [N_k-(t-1)] p(N_k) [1-p(N_k)]^{N_k-(t-1)-1}P(M_{N_k,t}=t-1), \nnb
\eee
and
\be
\lim_{N \to \infty} P(M_{N_k,t+1} = t) = \frac{\delta}{e^\delta} \cdot \gamma_t^{t-1} \triangleq \gamma_{t+1}^t. \nnb
\ee
Therefore, the limit $\lim\limits_{N \to \infty} P(M_{N_k,t+1} = m)$ exists for any $m \in \{0, 1,2,\cdots, t\}$, which completes the induction proof.
Thus, \eqref{equ:P-M-N-t=m-limit-exist} holds.

To prove \eqref{equ:limit-sum-is-1}, we can simply take limit in both sides of the following equality,
\be
\sum_{m=0}^{t-1} P(M_{N_k,t}=m) = 1, \forall t \in \{1,2,\cdots,D\}. \nnb
\ee

The proof is completed.

\subsection{Proof of Lemma~\ref{lem:an-bn-limsup}} \label{app:proof-of-lem-an-bn-limsup}
We prove this lemma by contradiction. Suppose that it is not true. Then,
since $\{a_N b_N\}$ is a bounded sequence, there exists a $\phi > 0$ such that
$\limsup_{N \to \infty} a_N b_N = \phi$,
which implies that there exists a subsequence $\{N_k\}_{k=1}^{\infty}$ such that
$\lim_{k \to \infty} a_{N_k} b_{N_k} = \phi > 0$.
Thus, there exists a $k_0 \ge 1$ such that
\be
a_{N_k} b_{N_k} > \frac{\phi}{2}, \quad \forall k \ge k_0. \nnb
\ee
Since $b_N \ge 0, \forall n$ and it is upper bounded, say $U>0$, we have
\be
a_{N_k} > \frac{\phi}{2U}, \quad \forall k \ge k_0, \nnb
\ee
which implies that
\be
\limsup_{k \to \infty} a_{N_k} \ge \frac{\phi}{2U} > 0. \nnb
\ee
Then we have
\be
\limsup_{N \to \infty} a_N \ge \limsup_{k \to \infty} a_{N_k} > 0, \nnb
\ee
which contradicts with $\limsup_{N \to \infty} a_N \le 0$.

Therefore, we must have $\limsup_{N \to \infty} a_N b_N \le 0$.

\subsection{Proof of Lemma~\ref{lem:p-1-N-to-1-e}} \label{app:proof-of-lem-p-1-N-to-1-e}
Again, we consider the first frame from slot 1 to slot $D$.
For any slot $t \in \{1,2,\cdots,D\}$, recall that we use random variable $M_{N,t}$ to
denote the number of finished stations  before slot $t$.
We then let random variable $S_{N,t} \in \{0,1\}$ denote the number of packets delivered in slot $t$.
Then, the probability of delivering a packet in slot $t$ is
\bee
P(S_{N,t}=1)   & = \sum\limits_{m=0}^{t-1} P(S_{N,t}=1|M_{N,t}=m) P(M_{N,t}=m)  \nnb \\
& = \sum\limits_{m=0}^{t-1} (N-m) p(N) \left[1- p(N) \right]^{N-m-1} P(M_{N,t}=m). \nnb
\eee
We consider the sequence $\{N_k\}=\{1,2,3,\cdots\}$.
Then $p(N)=1/N$ satisfies \eqref{equ:p(N)-to-0} and \eqref{equ:Np(N)-to-1} with $\delta=1$.
Thus, according to \eqref{equ:limit-is-delta-over-exp-delta} in Lemma~\ref{lem:a-key-lemma}
and \eqref{equ:P-M-N-t=m-limit-exist} and \eqref{equ:limit-sum-is-1} in Lemma~\ref{lem:prob-m-t-limit-exist},
we  get
\bee
& \lim_{N \to \infty} P(S_{N,t}=1)  = \lim\limits_{N \to \infty} \left\{ \sum\limits_{m=0}^{t-1} (N-m) p(N) \left[1- p(N) \right]^{N-m-1} P(M_{N,t}=m) \right\} \nnb \\
& =  \sum\limits_{m=0}^{t-1} \left\{\lim\limits_{N \to \infty} (N-m) p(N) \left[1- p(N) \right]^{N-m-1}  P(M_{N,t}=m)\right\} \nnb \\
& =  \sum\limits_{m=0}^{t-1} \left\{\lim\limits_{N \to \infty} (N-m) p(N) \left[1- p(N) \right]^{N-m-1}  \cdot \lim\limits_{N \to \infty} P(M_{N,t}=m)\right\} \nnb \\
& = \sum_{m=0}^{t-1} \frac{\delta}{e^\delta} \cdot \gamma_t^m   = \frac{1}{e^1} \cdot \sum_{m=0}^{t-1}  \gamma_t^m = \frac{1}{e} \cdot 1 = \frac{1}{e}. \nnb
\eee
Therefore, when $N$ goes to infinity, the probability of delivering a packet in any slot $t$ is $1/e$. Thus,
the system throughput also converges to $1/e$, i.e.,
\[
\lim_{N \to \infty} R(D, N, 1/N) = \lim_{N \to \infty} \frac{\sum_{t=1}^{D} P(S_{N,t}=1)}{D} =  1/e,
\]
which completes the proof.

\subsection{Proof of Lemma~\ref{lem:n-p-not-1-lead-to-suboptimal-rate}} \label{app:proof-of-lem-n-p-not-1-lead-to-suboptimal-rate}
For any slot $t \in \{1,2,\cdots,D\}$ in the first frame,
suppose that we have $M_{N_k,t}=m\in \{0, 1, \cdots, t-1\}$ finished stations  before slot $t$.
Clearly, there are $N_k-m$ active stations that have a packet at slot $t$.
In addition, based on Lemma ~\ref{lem:prob-m-t-limit-exist}, we have that
$\lim_{k \to \infty} P(M_{N_k,t}=m)=\gamma_t^m$ and
$\sum_{m=0}^{t-1} \gamma_t^m = 1$.

Recall that we use random variable $S_{N_k,t} \in \{0,1\}$ to denote the number of packets delivered in slot $t$.
Then, the probability of delivering a packet in slot $t$ is $P(S_{N_k,t}=1)$, and we have
\bee
& \lim_{k \to \infty} P(S_{N_k,t}=1)   = \lim_{k \to \infty}\sum_{m=0}^{t-1} P(S_{N_k,t}=1|M_{N_k,t}=m) P(M_{N_k,t}=m) \nnb \\
& = \sum_{m=0}^{t-1} \gamma_t^m \cdot \lim_{k \to \infty} (N_k-m)p(N_k) [1-p(N_k)]^{N_k-m-1}  = \sum_{m=0}^{t-1} \gamma_t^m \cdot  \frac{\delta}{e^\delta} = \frac{\delta}{e^\delta} < 1/e,
\eee
where the last inequality follows from the fact that
$\delta \neq 1$ and the fact that
function $f(x) = \frac{x}{e^{x}}$ is maximized
only when $x = 1$.
Therefore,
\bee
 \lim_{k \to \infty} R(D, N_k, p(N_k)) & = \lim_{k \to \infty} \frac{\sum_{t=1}^{D} P(S_{N,t}=1)}{D}   =  \frac{\delta}{e^{\delta}} < 1/e, \nnb
\eee
which completes the proof.

\subsection{Proof of Lemma~\ref{lem:opt-p-converge-to-0}} \label{app:proof-of-lem-opt-p-converge-to-0}

Since $p^*(D,N) > 0$, we have $\liminf\limits_{N \to \infty} p^*(D,N) \ge 0$.
Now we only need to show $\limsup\limits_{N \to \infty} p^*(D,N) = 0$.
Let us prove this by contradiction.
Suppose that
\be
\limsup_{N \to \infty} p^*(D,N) = \phi > 0. \label{equ:lem-opt-p-to-0-hypothesis} \nnb
\ee
This suggests that there exists a subsequence of sequence $\{N\}_{N=1}^{\infty}$, denoted by $\{N_k\}_{k=1}^{\infty}$,
such that
\be
\lim_{k \to \infty} p^*(D, N_k) = \phi. \nnb
\ee
This further shows that there exists a $k_0 \ge  1$ such that
\[
p^*(D, N_k) \ge \frac{\phi}{2} > 0, \quad \forall k \ge k_0.
\]

Let us consider $N_k$ stations where $N_k \ge \max \{N_{k_0}, \left \lceil \frac{2}{\phi} \right \rceil + D + 1\}$.
At any slot $t \in \{1,2,\cdots,D\}$, suppose that there are $M_{N,t}=m$ finished stations before slot $t$.
Clearly, $m \in \{0, 1, 2, \cdots, t-1\}$. The probability of delivering a packet in slot $t$ is
\be
(N_k-m) p^*(D, N_k) \left[ 1 - p^*(D, N_k) \right]^{N_k-m-1}. \nnb
\ee

We consider the following function
\be
f(p) = (N_k - m) p (1-p)^{N_k - m - 1},  \quad  p \in \left[\frac{\phi}{2}, 1 \right), \nnb
\ee
Then,
\bee
& \frac{df(p)}{dp}  = (N_k - m) (1-p)^{N_k - m -2} [1 - (N_k - m)p ]. \nnb
\eee

Since $N_k \ge   \max \{N_{k_0}, \left  \lceil \frac{2}{\phi} \right \rceil + D + 1\}$ and $p \in \left[\frac{\phi}{2}, 1 \right)$,
we have $N_k-m\geq N_k-D>0$ and
\[
1-(N_k-m)p\leq 1-(N_k-D)\frac{\phi}2 \leq 1 - \left( \left \lceil \frac{2}{\phi} \right \rceil+1\right) \frac{\phi}{2} \le
- \frac{\phi}{2} < 0.
\]
Thus, for $\phi/2\leq p < 1$,
\be
\frac{df(p)}{dp} < 0. \nnb
\ee
In addition, since  $N_k \ge N_{k_0}$, we have $p^*(D, N_k) \ge \phi/2, \forall k \ge k_0$.
Then, we have
\bee
 0  & \le f(p^*(D, N_k))  = (N_k -m ) p^*(D, N_k) \left[ 1 - p^*(D, N_k) \right]^{N_k-m-1} \nnb \\
& \le f(\phi/2) = (N_k - m) \frac{\phi}{2} \left(1- \frac{\phi}{2}\right)^{N_k - m - 1},  \forall k \ge k_0
\label{equ:squeeze-f(p)-inequ}
\eee
Note that
\be
\lim_{k \to \infty} (N_k - m) \frac{\phi}{2} \left(1- \frac{\phi}{2}\right)^{N_k - m - 1} = 0. \nnb
\ee
Thus, by applying the squeeze theorem to \eqref{equ:squeeze-f(p)-inequ}, we have
\be
\lim_{k \to \infty} (N_k -m) p^*(D, N_k) \left[ 1 - p^*(D, N_k) \right]^{N_k-m-1}  = 0. \nnb
\ee
Therefore, the probability of delivering a packet in slot $t$ converges as
\bee
& \lim_{k \to \infty} P(S_{N_k,t}=1)  = \lim_{k \to \infty}  \sum_{m=0}^{t-1} P(S_{N_k,t}=1|M_{N_k,t}=m) P(M_{N_k,t}=m) \nnb \\
&= \lim\limits_{k \to \infty} \sum\limits_{m=0}^{t-1} (N_k-m) p^*(D, N_k) \left[ 1 - p^*(D, N_k) \right]^{N_k-m-1} P(M_{N_k, t}=m) \nnb \\
& = \sum\limits_{m=0}^{t-1} \lim\limits_{k \to \infty} (N_k-m) p^*(D, N_k) \left[ 1 - p^*(D, N_k) \right]^{N_k-m-1} P(M_{N_k, t}=m)  = \sum_{m=0}^{t-1} 0 = 0, \nnb
\eee
where the second last equality follows from the fact that if $\lim_{n \to \infty} a_n = 0$ and $\{b_n \}$ is bounded (here we have $ |P(M_{N_k,t}, = m)| \le 1$),
then $\lim_{n \to \infty} a_n b_n = 0$.
Thus, in any slot of a frame, the probability of delivering a packet converges to 0
(under the subsequence $\{N_k\}$), yielding to
\be
\lim_{k \to \infty} R^*(D, N_k) = \lim_{k \to \infty} \frac{\sum_{t=1}^{D} P(S_{N_k,t}=1)}{D} = 0, \nnb
\ee
which contradicts to Theorem~\ref{thm:converge-to-1-over-e}.
This completes the proof.

\subsection{Proof of Theorem~\ref{thm:opt-R-converge-to-1-over-e-dynamic}} \label{app:proof-of-thm-opt-R-converge-to-1-over-e-dynamic}
By following the similar induction steps in the proof of Lemma~\ref{lem:prob-m-t-limit-exist},
it is straightforward to show that \eqref{equ:P-M-N-t=m-limit-exist} and \eqref{equ:limit-sum-is-1} hold
under the optimal retransmission policy $\bm{p}^*=\{p^*_{\textsf{dynamic}}(n,t), n=1,2,\cdots,N, t=1,2,\cdots, D \}$.

Then, according to \eqref{equ:algo-R-based-on-S-t}, we have
\[
\lim_{N \to \infty} R_{\textsf{dynamic}}(D,N, \bm{p}^*)   = \frac{ \sum_{t=1}^D \lim_{N\to \infty} P(S_t=1) }{D}.
\]
In addition,
\bee
& \lim_{N\to \infty} P(S_t=1)  =  \lim\limits_{N\to \infty} \sum\limits_{m=0}^{t-1} P(S_t=1|M_{N,t-1}=m) P(M_{N,t-1}=m) \nnb \\
& =   \lim\limits_{N\to \infty} \sum\limits_{m=0}^{t-1} (N-m) p_{\textsf{dynamic}}^*(N-m, t) \left[1 - p_{\textsf{dynamic}}^*(N-m,t)\right]^{N-m-1}  P(M_{N,t}=m) \nnb \\
& =  \lim\limits_{N\to \infty} \sum\limits_{m=0}^{t-1} (N-m) \frac{1}{N-m} \left(1 - \frac{1}{N-m}\right)^{N-m-1}  P(M_{N,t}=m) \nnb \\
& = \sum_{m=0}^{t-1}  \lim\limits_{N\to \infty}  \left(1 - \frac{1}{N-m}\right)^{N-m-1} \cdot \lim\limits_{N \to \infty} P(M_{N,t}=m)
 = \sum_{m=0}^{t-1}  \frac{1}{e} \cdot \gamma_t^m  = \frac{1}{e} \cdot \sum_{m=0}^{t-1}   \gamma_t^m  = \frac{1}{e} \cdot 1 = \frac{1}{e}. \nnb
\eee
Thus,
\[
\lim_{N \to \infty} R_{\textsf{dynamic}}(D,N, \bm{p}^*)   = \frac{ \sum\limits_{t=1}^D \lim\limits_{N\to \infty} P(S_t=1) }{D} = \frac{\sum\limits_{t=1}^D \frac{1}{e}}{D} = \frac{1}{e}.
\]

\ifx \ISTR \undefined
\else
\fi

\end{document}